\documentclass[
    a4paper,
    prb,
    bibnotes, 
    twocolumn,
    superscriptaddress,
    floatfix
]{revtex4-2}
\usepackage{amssymb,amsmath}
\usepackage{amsmath}
\usepackage{graphicx}% Include figure files
\usepackage{dcolumn}% Align table columns on decimal point
\usepackage{bm}% bold math
\usepackage{color,xcolor}
\usepackage{multirow}
\usepackage{verbatim}
\usepackage[colorlinks,anchorcolor=blue,urlcolor=blue,citecolor=blue]{hyperref}

\newcommand{\Vq}{{\bm{q}}}
\newcommand{\VP}{{\bm{p}}}
\newcommand{\VK}{{\bm{k}}}
\newcommand{\Vu}{{\bm{u}}}

\newcommand{\Vv}{{\bm{v}}}
\newcommand{\Vkap}{{\bm{\kappa}}}
\newcommand{\Vphi}{{\bm{\Phi}}}
\newcommand{\tu}{\tilde{u}}
\newcommand{\fneq}{f_N^{}}
\newcommand{\freq}{f_R^{}}

\newcommand{\AV}[1]{\langle #1 \rangle}
\makeatletter

\newcommand{\Rmnum}[1]{\expandafter\@slowromancap\romannumeral #1@}
\makeatother

\begin{document}
\title{A unified theory of second sound in two dimensional materials}
\author{Man-Yu Shang}
\affiliation{School of Physics and Wuhan National High Magnetic Field Center, Huazhong University of Science and Technology, 430074 Wuhan, P. R. China}
\author{Wen-Hao Mao}
\affiliation{School of Physics and Wuhan National High Magnetic Field Center, Huazhong University of Science and Technology, 430074 Wuhan, P. R. China}
\author{Nuo Yang}
\affiliation{School of Energy and Power Engineering, \\Huazhong University of Science and Technology, 430074 Wuhan, P. R. China}
\author{Baowen Li}
\email{baowen.li@colorado.edu}
\affiliation{Paul M. Rady Department of Mechanical Engineering, University of Colorado, Boulder, CO, 80309, USA}
\affiliation{Department of Physics, University of Colorado, Boulder, CO, 80309, USA}
\author{Jing-Tao L\"u}
\email{jtlu@hust.edu.cn}
\affiliation{School of Physics and Wuhan National High Magnetic Field Center, Huazhong University of Science and Technology, 430074 Wuhan, P. R. China}

\begin{abstract}
%Second sound refers to wave-like propagation of temperature field in dielectric solids, emerging from frequent anharmonic phonon scattering. 
%Two-dimensional (2D) materials are predicted to host second sound in a much wider temperature range than their three-dimensional counterparts due to presence of flexual phonons. However, how their quadratic dispersion influences the emergence of second sound is still poorly understood, leading to different expressions for the sound velocity.

%Two-dimensional (2D) materials are predicted to host second sound in a much wider temperature range than their three-dimensional counterparts due to presence of flexual phonons. However, its experimental observation remains elusive. 
%Here, 
We develop a unified theory for the second sound in two dimensional materials.  Previously studied drifting and driftless second sound are two limiting cases of the theory, corresponding to the drift and diffusive part of the energy flux, respectively. We find that due to the presence of quadratic flexural phonons the drifting second sound does not exist in the thermodynamic limit, while the driftless mode is less affected. This is understood as a result of infinite effective inertia of flexual phonons, due to their constant density states and divergent Bose-Einstein distribution in the long wave length limit. Consequently, the group velocity of the drifting mode is smaller than that of the driftless mode. However, upon tensile strain, the velocity of drifting mode becomes larger. Both of them increase with tensile strain due to the linearization of the flexural phonon dispersion. Our results clarify several puzzles encountered previously and pave the way for exploring wave-like heat transport beyond hydrodynamic regime. %Finally, possible connection with anomalous thermal conductivity is discussed.
%fully numerical study and experimental search of second sound in 2D materials, which remains elusive despite recent intense study.

%1st submission
%Phonon second sound has been predicted to exist in a much wider temperature range in two-dimensional (2D) materials than their three-dimensional counterpart. However, due to the presence of flexural phonons with quadratic dispersion, how second sound emerges from microscopic phonon dynamics in 2D system is poorly understood. This has led to different expressions for the second sound velocity in previous studies.
%Here, we develop a unified theory to resolve this controversy. Previously studied drifting and driftless second sound are two limiting cases of our unified theory. Surprisingly, we find that quadratic flexural phonons do not support drifting second sound in the thermodynamic limit, and severely reduce its group velocity for finite size sample. Meanwhile, the driftless sound is less affected. We explain this finding as a result of infinite effective mass density of the flexural phonons. Our results pave the way for fully numerical study and experimental search of second sound in 2D materials, which remains elusive despite recent intense study.
\end{abstract}

\maketitle

\section{Introduction}
\begin{figure}
	\centering 
	\includegraphics[scale=0.45]{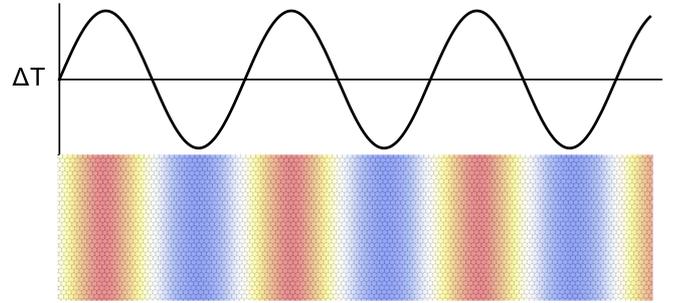}
	\caption{Schematic diagram of the second sound propagation.} 
	\label{fig:ss-sketch} 
\end{figure}

While the diffusive Fourier heat conduction is ubiquitous in bulk solids, the violation in various circumstances, in particular in low dimensional systems, has been observed and is becoming a strong focus of current research in condensed matter and statistical physics, nano-material science and engineering\cite{Chen2021,Wang2008,Dhar2008,Nianbei-RMP,gu_phononic_2018,lepri_thermal_2003}. One example of such violation is the wave-like propagation of temperature field (Fig.~\ref{fig:ss-sketch}), termed second sound, an emergent many-body phenomenon resulting from frequent phonon scattering
\cite{beck_phonon_1974,joseph_heat_1989-1,Lee-Review}.
%Distinct from ubiquitous Fourier heat diffusion, it 
Thermal wave transport provides new opportunities for heat management, information processing and novel device applications\cite{Nianbei-RMP,gu_phononic_2018,wang2020,Nakamura2019,li_transforming_2021}.  Exploring its role in anomalous thermal transport may offer new insight into the divergent thermal conductivity of low dimensional system\cite{lepri_thermal_2003,Dhar2008,Nianbei-RMP,gu_phononic_2018}.
Early research has led to observation of second sound only in a handful of materials\cite{ward_velocity_1951,ward_iii_1952,sussmann_thermal_1963,gurzhi_hydrodynamic_1968,guyer_solution_1966,guyer_thermal_1966,hardy_phonon_1970,ackerman_second_1966,mcnelly_heat_1970,narayanamurti_observation_1972,koreeda_second_2007}.
%Soon after its prediction and observation in He, the analysis was extended to dielectric solids%
%, where phonons are the dominant energy carriers
%\cite{ward_iii_1952,sussmann_thermal_1963,gurzhi_hydrodynamic_1968,guyer_solution_1966,guyer_thermal_1966,hardy_phonon_1970}. But its observation was limited to only a handful of materials\cite{ackerman_second_1966,mcnelly_heat_1970,narayanamurti_observation_1972,koreeda_second_2007}. 
Recent progress on phonon\cite{lee_hydrodynamic_2015,cepellotti_phonon_2015,ding_phonon_2017,martelli_thermal_2018,machida_observation_2018,machida_phonon_2020,huberman_observation_2019,cepellotti_thermal_2016,shang_heat_2020,guo_heat_2017,lee_hydrodynamic_2017,luo_direct_2019,beardo_phonon_2020,torres_emergence_2018,zhang2020violation,yu_perspective_2021} and electron\cite{crossno_observation_2016,bandurin_negative_2016,moll_evidence_2016,sulpizio_visualizing_2019,gallagher_quantum-critical_2019,ella_simultaneous_2019,berdyugin_measuring_2019} hydrodynamic transport in two dimensional (2D) materials has triggered its renewed interest, based on which coupled electron-phonon hydrodynamics has been anticipated\cite{levchenko_transport_2020,huang_electron-phonon_2021,narozhny_electronic_2019}. However, direct experimental observation of phonon second sound and its connection with anomalous thermal conductivity in 2D materials are still lacking.

Theoretical analysis has identified two types of second sound, denoted as drifting and driftless modes, respectively\cite{hardy_phonon_1970}. The drifting mode exists in the hydrodynamic transport regime, where crystal momentum conservation is approximately fulfilled during phonon scattering. This requires the momentum-conserving normal scattering  ($N$-scattering)  process dominates over the non-conserving  processes ($R$-scattering). The latter includes Umklapp scattering, impurity scattering, and scattering with other quasi-particles.  
The existence of driftless mode requires that the heat-carrying phonons have similar relaxation time, which should be much longer than the inverse of external driving frequency. It existence does not rely on the hydrodynamic conditions and is possible even in the diffusive regime. Although these two types of second sound has been noticed long time ago\cite{hardy_phonon_1970,hardy_velocity_1971}, their different nature has not been clarified, leaving the experimentally observed second sound in different materials unclassified\cite{huberman_observation_2019,beardo_observation_2021}. Moreover, in the two seminal works on 2D materials\cite{cepellotti_phonon_2015,lee_hydrodynamic_2015}, to avoid an infrared divergence introduced by the quadratic flexural phonons, two different expressions for the second sound velocity have been used, which needs further clarification. 

By combining momentum and energy flux balance equations, we develop a unified theory to understand the nature of phonon second sound in 2D materials.
We show that the drifting and driftless modes emerge in our theory as two limiting cases, and they are rooted in drift and diffusive part of the total energy flux, corresponding to the 1st and 2nd term at the right hand side (rhs) of Eq.~(\ref{eq:qm0}), respectively. 
%Both of them contribute to the resulting second sound, with the dominant contribution from the former at low frequency and from the latter at high frequency. % 
%Specifically, in 2D materials, the coexistence of phonons with linear and quadratic dispersion results in dynamics with mixed relativistic and non-relativistic characteristics.
More importantly, in ideal non-strained 2D materials, the constant density of states of quadratic flexural phonons and divergent Bose-Einstein distribution in the long wave length limit together give rise to logarithmic divergence of the phonon number density with system size. Consequently, the drifting second sound does not exist in the thermodynamic limit. This is a common feature of bosonic quasi-particles with quadratic dispersion but without number conservation, and reveals the physical origin of the infrared divergence encountered in previous works\cite{simoncelli_generalization_2020,cepellotti_phonon_2015,lee_hydrodynamic_2015,lee_hydrodynamic_2017,shang_heat_2020}.

\section{Theory}
\subsection{Balance equations}
We follow a kinetic approach and employ the Peierls-Boltzmann equation in the Callaway approximation to describe the phonon transport\cite{callaway_model_1959},
\begin{equation}
\frac{\partial{f}_{i\VK}}{\partial{t}}+\Vv_{i\VK}\cdot\nabla{f}_{i\VK}
=-\frac{f_{i\VK}-f_{R,i\VK}^{}}{\tau_{R,i\VK}}-\frac{f_{i\VK}-f_{N,i\VK}^{}}{\tau_{N,i\VK}}.
\label{eq:callaway}
\end{equation}
Here, $\tau_{R,i\VK}$ and $\tau_{N,i\VK}$ are two mode-resolved, $\VK$-dependent relaxation time are introduced to describe $R$ and $N$ processes, respectively. $f_{i\VK}$ is the nonequilibrium phonon distribution, $\Vv_{i\VK}$ is the phonon group velocity, and $i$ is the phonon branch index. Hereafter, we omit $i\VK$ for brevity when there is no ambiguity. It is known that with $R$-scattering the system relaxes to the Bose-Einstein distribution 
\begin{align}
f_{R,i\VK} = \{{\rm exp}[\beta\hbar\omega_{i\VK}]-1\}^{-1}, 
\end{align}
where $\hbar$ is the reduced Planck constant, $\omega_{i\VK}$ is the phonon angular frequency, $\beta=(k_BT)^{-1}$ is the inverse temperature with the Boltzmann constant $k_B$ and the absolute temperature $T$. Meanwhile, in the presence of only $N$-scattering, the system relaxes instead to a drifted Bose-Einstein distribution 
\begin{align}
f_{N,i\VK}=\{{\rm exp}[\beta(\hbar\omega_{i\VK}-\hbar\VK\cdot\Vu)]-1\}^{-1} 
\end{align}
with a characteristic drift velocity $\Vu$ common to all phonon modes.

The balance equations can then be derived by considering the (quasi-)conserved quantities in the kinetics. Since phonons represent thermal excitation of the atomic motion, their number is not a conserved quantity. Thus, we are left only with energy and crystal momentum,
\begin{align}
{\partial_t{E}}+\nabla\cdot\Vq&=0, \label{eq:econ}\\
\partial_t{\VP}+\nabla\cdot\Vphi&=-\AV{\tau^{-1}_{R}}_p\VP, 
\label{eq:pcon}
\end{align}
$E$ and $\VP$ the energy and momentum density, and 
$\Vq$ and $\Vphi$ the corresponding fluxes.
The averaged inverse relaxation time $\AV{\tau_R^{-1}}_p$ [Eq.~(\ref{eq:taurp})] characterizes the relaxation of $\VP$ due to $R$-scattering.
The energy conservation results from ignoring scattering processes that transfer energy to other quasi-particles, i.e., electrons. 
We also need an equation for the energy flux $\Vq$
\begin{align}
    \AV{\tau_{c}}_q \partial_t \Vq + \Vq = \chi W^{(0)}\Vu - \bm{\kappa} \cdot \nabla T.
    \label{eq:qm0}
\end{align}
The total energy flux includes two contributions. The first term at the rhs is due to the collective phonon drift motion, while the second is due to a temperature gradient. 
Here, $\AV{\tau_c}_q$ [Eq.~(\ref{eq:qtau})] is a characteristic relaxation time of $\Vq$, $\kappa$ [Eq.~(\ref{eq:kappa})] is the thermal conductivity in the relaxation time approximation,  $W^{(0)}$ [Eq.~(\ref{eq:defw})] is the enthalpy function evaluated by approximating $f_{i\VK}\approx f^{(0)}_{i\VK} = f_{N,i\VK}$, and $\chi$ [Eq.~(\ref{eq:eq0})] is an averaged dimensionless parameter characterizing the relative contribution of $N$-scattering to the total scattering rate. Details of the derivation and the definition of these variables can be found in App.~\ref{app:theory}. 
%We note that Callaway has written the second term at the right side of Eq.~(\ref{eq:qm0}) into a form similar to the first, such that  a total thermal conductivity, larger than $\kappa$, can then be defined. 

\subsection{Linear phonons}
We start by considering a single phonon branch with linear dispersion, where the energy flux and the momentum density are simply proportional to each other, i.e.,  $\Vq=v_g^2 \VP$. To linear order in $u$, we have $\Vq = W^{(0)} \Vu$.  This holds for `relativistic' quasi-particles with linear dispersion. Using this equivalence, a Guyer-Krumhansl equation can be  derived\cite{guyer_solution_1966,guyer_thermal_1966,guo_phonon_2015}. 
Combining with Eq.~(\ref{eq:econ}), we can get a wave solution for the temperature field, 
with group velocity $v = v_g/\sqrt{D}$, where $D$ is the system dimension, $v_g$ is the phonon group velocity (see also App. \ref{app:helium-ss}). 

We can use this simple case to make connection with second sound in helium \Rmnum{2} (App. \ref{app:helium-ss}). The common drift velocity $\Vu$ here plays the role of relative velocity between the normal and super fluid in helium II. Both of them sustain even in the absence of external driving, carry no entropy and are essential for propagation of second sound. However, they are from different microscopic origin. Here, it requires frequent momentum-conserving $N$-scattering to sustain the collective drift motion, while in helium it relies on Bose-Einstein condensation to produce super fluid helium and its relative motion with the normal fluid.

In the presence of more phonon branches, i.e., longitudinal and transverse acoustic branches, having different group velocity $v_{g,i}$,  the proportionality between $\VP$ and $\Vq$ does not hold,
giving rise to the drifting and driftless second sound with their velocity\cite{hardy_velocity_1971}
\begin{align}
v_{p} \approx \sqrt{\left(\sum_i v_{g,i}^{-D}\right)/\left(D\sum_i v_{g,i}^{-(D+2)}\right)}
\end{align}
and 
\begin{align}
v_{q} \approx \sqrt{\sum_i v_{g,i}^{2-D}/\left(D\sum_i v_{g,i}^{-D}\right)},
\end{align}
respectively\footnote{These expressions hold in the constant relaxation time approximation, i.e., $\AV{\tau_c}_\kappa = \AV{\tau_c}_q=\AV{\tau_c}$.}.
The situation changes qualitatively in the presence of flexural phonons with quadratic dispersion.

\subsection{Flexural phonons}
In the ideal, non-strained 2D system, the flexural phonons have quadratic dispersion with constant density of states. The large Gr\"uneisen parameter indicates their dominant role in the anharmonic $N$-scattering. This has been attributed to the physical mechanism leading to phonon hydrodynamics in much wider temperature range in 2D materials\cite{lee_hydrodynamic_2015}.
However, the quadratic distribution poses difficulties in the traditional kinetic theory treatment of phonon hydrodynamics, where a small-$u$ expansion on $f_{N,i\VK}$ is performed, i.e., 
\begin{align}
f_{N} \approx f_{R} + \beta f_R(f_R+1)\hbar \VK\cdot \Vu.
\end{align}
Unfortunately, this expansion fails for $\hbar\omega_{i\VK} \sim \hbar \VK\cdot \Vu$, which is always true for a quadratic dispersion in the long wavelength limit. More severe is the unphysical case when $\hbar\omega_{\VK} < \hbar\VK\cdot\Vu$ and $f_{\VK}<0$. 
This has been the main obstacle in understanding 2D phonon hydrodynamics and in applying fully numerical approach to realistic materials\cite{simoncelli_generalization_2020}. 

To focus on this problem, we postpone the full analysis and consider the flexural phonon branch only. We derive the hydrodynamic equation using the full form of $\fneq$ instead. We furthermore introduce an effective mass $m^* = \hbar/2a$ for the flexural phonons, such that  
\begin{align}
    \hbar\omega_k = \hbar a k^2 = \hbar^2 k^2/2m^*.
\end{align}
Note that $m^*$ is introduced for notational convenience and is not the atomic mass.  
We can then derive a generalized Euler equation from Eq.~(\ref{eq:pcon}) (App.~\ref{app:theory})  
\begin{align}
(\partial_t + \Vu\cdot \nabla) \Vu + (\nabla\cdot \Vu)\Vu +  \nabla P^{(0)}/\rho_N^{(0)} &= -\AV{\tau_{R}^{-1}}_p\Vu.
\label{eq:euler1}
\end{align}
Here, $P^{(0)}$ is the effective pressure of the phonon gas [Eq.~(\ref{eq:p0n})], $\rho_N^{(0)} = n_N^{(0)}m^*$ is the effective mass density, with the phonon number density $n_N^{(0)}=L^{-2}\sum_{\VK} f_{N,\VK}$. With these effective parameters, Eq.~(\ref{eq:euler1}) takes the standard form for non-relativistic particles\cite{LandauBook6}.
%, with two differences: (1) the phonon number is not conserved, (2) the zeroth order quantities with superscript $(0)$ are evaluated with $f_N$, instead of $f_R$. As a result of (2), the Galilean invariance is not fulfilled.

To consider wave solutions, we ignore terms nonlinear in $\Vu$. 
One important feature of flexural phonons is that $\rho^{(0)}_N$ diverges logarithmically with system size $L$
\begin{equation}
\rho_N^{(0)}(u=0) \propto {\rm ln}L.\label{eq:rho0limit}
\end{equation}
This is due to their constant density of states at $k=0$ where $f_{R,\VK}$ diverges. The divergent $\rho_N^{(0)}$ results in an equation about $\Vu$ as $\partial_t \Vu =-\AV{\tau^{-1}_{R}}_p\Vu$, with the steady-state solution $\Vu=0$. Thus, we reach one important result: the quadratic flexural phonons do not support drifting second sound in the thermodynamic limit. Consequently, the problem with negative occupation $f_\VK <0$ does not occur. This is a general feature of 2D bosonic quasi-particles with quadratic dispersion that lack number conservation. We have provided an intuitive explanation of this result as a consequence of their infinite effective inertia. 

%We note that, in practice, several factors can lead to a finite $\rho_N^{(0)}$ and consequently a slow drifting second sound. Firstly, the finite size of the sample introduces a low cutoff to the wave vector
%$k_{\rm cut} \sim 2\pi/L$, where $L$ is the length of the 2D sample.  Secondly, it is known that the low frequency flexural mode in 2D materials is strongly anharmonic, which may lead to a renormalized dispersion $\omega_k \propto k^\gamma$ with $1<\gamma <2$\cite{Mariani_flexural_2008} (see however Ref.~\onlinecite{aseginolaza_bending_2020} for an opposite view), removing the divergence in $n_N^{(0)}$. Thirdly, tensile strain can harden the flexural mode and introduce a linear dispersion near $k=0$. Importantly, although in practice the divergence can be avoided,  $\rho^{(0)}_N$ is still much larger than $\rho^{(0)}_L$ (Fig. C3), with $\rho_L^{(0)} \equiv v_g^{-2} W_L^{(0)}$ the effective mass density for the linear mode. This means the drifting second sound velocity can be drastically reduced by the presence of flexural mode, as confirmed in the following numerical result (Fig.~\ref{fig:w-k}). 

Considering instead the energy flux, we obtain a damped wave solution 
for the driftless second sound with velocity
\begin{align}
v_q\approx \sqrt{\kappa/(C_N^{(0)}\AV{\tau_c}_q)}.
\end{align}
It depends on the heat capacity of flexural phonons $C_N^{(0)}$, instead of the divergent $\rho^{(0)}_N$.  

The existence of driftless sound mode can be understood as follows. When a time dependent external temperature gradient is applied to the system, the energy current response is also time dependent. The finite response time of the system is taken into account by the first term in Eq.~(\ref{eq:qm0}). In the case  $\AV{\tau_c}_q\partial_t\Vq \gg \Vq$,  Eqs.~(\ref{eq:econ}) and (\ref{eq:qm0}) allow damped wave solutions. This situation is similar to the optical response of free electrons in the Drude model. We get a frequency-dependent thermal conductivity\cite{guyer_solution_1966,volz_thermal_2001,chaput_direct_2013,hua_space_2020,koh_frequency_2007}
\begin{align}
    \kappa(\omega) = \frac{\kappa}{1-i\omega \AV{\tau_c}_q}.
\end{align}
The real part represents in-phase response of $\Vq$ to the time dependent temperature field, resulting in dissipation, while the imaginary part has a $\pi/2$ phase lag and gives rise to wave propagation. It becomes dominant for $\omega \gg \AV{\tau_c}_q^{-1}$. This means that the existence of driftless second sound does not rely on the stringent phonon hydrodynamic conditions, but requires a high frequency excitation. Recent experimental observation of second sound under high frequency excitation in Ge seems to fall into this regime\cite{beardo_observation_2021}.

\begin{figure*}
	\centering 
	\includegraphics[scale=0.8]{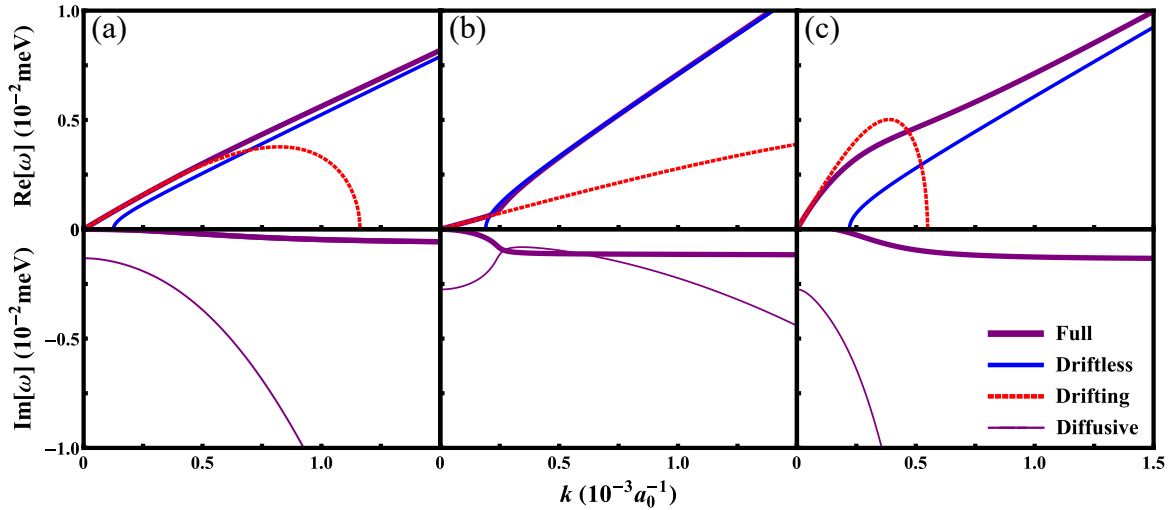} 
	\caption{The second sound dispersion relation of graphene calculated using  parameters obtained from DFT calculations at $T=100$ K shown in Figs.~(\ref{fig:wk-com}-\ref{fig:rho-all}). We take $L = 50 \mu$m, corresponding to a low cutoff wave vector $k_{\rm cut}\sim 3 \times 10^{-5} a_0^{-1}$, where $a_0=2.47$ \AA~is the lattice constant of graphene.
	The red dashed and blue solid lines correspond to solutions of the drifting  and driftless modes, while the purple lines are the full solutions [Eq.~(\ref{eq:disg})]. The drifting modes are calculated by taking $\Vq = W^{(0)} \Vu$, while the driftless modes are obtained by taking $\Vu=0$ in Eq.~(\ref{eq:disg}). The upper and lower panel give the real and imaginary parts, respectively. Only positive solutions are shown in the upper panel, while only the full solutions are shown in the lower. In the lower panel, the absolute value of the thick line gives the lifetime of the second sound in the upper panel, while the thin line is the diffusive mode with a zero real part. Thus, the actual heat transport composes both the sound and the diffusive mode simultaneously. In (a) the force constants from DFT calculations are used directly, while in (b) additional rotational symmetry is applied to the force constants. In (c) the flexural phonons are excluded. 		} 
	\label{fig:w-k} 
\end{figure*}

\subsection{Full analysis}
We now turn to  the full form of the balance equations by including two linear and one flexural acoustic branches. We furthermore include the viscous dissipation, which generates damping of the second sound. Correspondingly, the generalized Euler equation is modified to a Navier-Stokes equation (App.~\ref{app:theory} of the Appendix)
\begin{align}
(\partial_t + \Vu\cdot \nabla) \Vu &+ (\nabla\cdot \Vu)\Vu   = -\nabla P^{(0)}/\rho^{(0)}\\
&+\eta /\rho^{(0)} \nabla^2 \Vu+\xi /\rho^{(0)} \nabla(\nabla\cdot\Vu)-\AV{\tau^{-1}_{R}}_p\Vu.\nonumber
\label{eq:ns1}
\end{align}
The bulk ($\eta$) and shear ($\xi$) viscosity coefficients describes the hydrodynamic dissipation, with $\xi=0$ in 2D (App.~\ref{app:transport}), and $\rho^{(0)}$, $P^{(0)}$ include contributions from all phonon branches.
We note that all the zeroth order quantities are evaluated using $f_N$, instead of the thermal equilibrium $f_R$. This is different from the standard relativistic and non-relativistic hydrodynamics. Consequently, phonons do not fulfill the Lorentz or Galilean invariance\cite{narozhny_electronic_2019}. Only when we consider second sound propagation and keep only terms linear in $u$, can they be recovered (App.~\ref{subsec:renonrel}).  
In that case, considering small deviations of $\Vu$, $T$ and $\Vq$  on top of their equilibrium value with the form
\begin{align}
     \Vu, \delta T, \delta \Vq \propto {\rm exp}(-i\omega t + i \VK\cdot {\bf r}),
\end{align}
we obtain a set of linear equations from Eqs.~(\ref{eq:econ},\ref{eq:qm0},\ref{eq:ns1})
\begin{widetext}
\begin{align}
    \left(
    \begin{array}{ccc}
        \omega  &  0 & -k_\alpha/C^{(0)} \\
        -k_\alpha C^{(0)}_P/\rho^{(0)} & \omega + i \AV{\tau_R^{-1}}_p +i (\eta+\xi)k_\alpha^2/\rho^{(0)} & 0\\
        i k_\alpha \kappa & -\chi W^{(0)} & 1-i\omega\AV{\tau_c}_q \\
    \end{array}
    \right)
    \left(
    \begin{array}{ccc}
         \delta T\\
         u_\alpha \\
         \delta q_\alpha
    \end{array}
    \right)
    =
    0,
    \label{eq:disg}
\end{align}
\end{widetext}
where $C^{(0)}_P = \partial P^{(0)}/\partial T$, $\alpha=x, y, z$.
Dispersion relations of the associated modes can be obtained from the condition ${\rm det} A = 0$, where $A$ is the $3\times 3$ matrix in Eq.~(\ref{eq:disg}). 

Before presenting the numerical result, we can show that the drifting and driftless modes show up as two limiting cases of Eq.~(\ref{eq:disg}). 
In the first limit, when the drifting part dominates, the energy flux $\Vq \approx \Vq^{(0)}=W^{(0)}\Vu$. Equation~(\ref{eq:disg}) gives
\begin{align}
    \omega_{p}(k) \approx &\pm \sqrt{\frac{W^{(0)}(\sum_i \alpha_i C^{(0)}_{i})}{\rho^{(0)}C^{(0)}} k^2-\frac{1}{4}\delta^2(k)}  -\frac{i}{2}\delta(k),
    \label{eq:mix}
\end{align}
with
$\delta(k) =\AV{\tau_{R}^{-1}}_p+ (\eta+\xi)k^2/\rho^{(0)}$ (red dashed lines in Fig.~\ref{fig:w-k}).
The coefficient $\alpha_L=1/2$ for linear mode and $\alpha_N=1$ for quadratic flexural mode. $\rho^{(0)}$ in the denominator indicates their origin from the momentum balance equation. Since $\rho^{(0)}_N$ diverges as ${\rm ln}L$  in the thermodynamic limit, there is no drifting second sound solution. 
In the other limit, when the angular frequency is much larger than the inverse decay time of the energy flux, $\omega \gg \AV{\tau_c}_q^{-1}$, there can still be wave solution even when $u = 0$. We get the driftless sound mode
\begin{align}
    \omega_{q}(k) &= \pm\sqrt{\frac{\kappa}{C^{(0)}\AV{\tau^{}_c}_q}k^2-\frac{\AV{\tau_c^{}}_q^{-2}}{4}}  - \frac{i}{2} \AV{\tau_c^{}}_q^{-1},
    \label{eq:driftless}
\end{align}
which are mainly associated with the second term at the rhs of Eq.~(\ref{eq:qm0}) (blue solid lines in Fig.~\ref{fig:w-k}). 

We note that, in practice, several factors can lead to a finite $\rho_N^{(0)}$ and consequently a slow drifting second sound. Firstly, the finite size of the sample introduces a low cutoff to the wave vector
$k_{\rm cut} \sim 2\pi/L$, where $L$ is the length of the 2D sample.  Secondly, it is known that the low frequency flexural mode in 2D materials is strongly anharmonic, which may lead to a renormalized dispersion $\omega_k \propto k^\gamma$ with $1<\gamma <2$\cite{Mariani_flexural_2008} (see however Ref.~\onlinecite{aseginolaza_bending_2020} for an opposite view), removing the divergence in $n_N^{(0)}$. Thirdly, tensile strain can harden the flexural mode and introduce a linear dispersion near $k=0$. Notably, although in practice the divergence can be avoided, the drifting second sound velocity can be drastically reduced by the quadratic dispersion of flexural mode (Fig.~\ref{fig:w-k}). 

%with $\rho_L^{(0)} \equiv v_g^{-2} W_L^{(0)}$ the effective mass density for the linear mode

\section{Numerical results}
We now turn to fully numerical calculations. Figure~\ref{fig:w-k} presents the dispersion relation obtained by solving Eq.~(\ref{eq:disg}) numerically, where only the positive solution is shown (purple solid lines). We use realistic parameters obtained from density functional theory (DFT) calculations of graphene  (details in App.~\ref{app:dft}). To study the drifting mode, we have chosen a cut off wave vector corresponding to finite size system with $L=50$ $\mu$m. Similar results are obtained for single layer Boron Nitride (not shown here).
%(Fig.~\ref{fig:wk-com-BN}-\ref{fig:vepsilon-BN}). 

The drifting mode (red dashed) exists in the long wave length limit with the upper cut off determined by the condition $\omega_p(k) \approx \delta (k)/2$ [Eq.~(\ref{eq:mix})]. On the other hand, the driftless mode (blue solid) has a lower cut off determined by $\omega_q(k) \approx \AV{\tau_c}_q^{-1}/2$ [Eq.~(\ref{eq:driftless})]. These two limiting wave vectors determine the overlap regime of the two types of second sound. The ideal quadratic dispersion of flexural phonons is not guaranteed from the numerical force constants. Additional symmetrization is applied to recover the quadratic dispersion \cite{Jes2016Physically}. Figure~\ref{fig:w-k} (a) and (b) show the resulting second sound dispersion before and after the symmetrization. A clear transition from drifting to driftless mode is observed in Fig.~\ref{fig:w-k} (b), but is difficult to see in  Fig.~\ref{fig:w-k} (a). This shows a small deviation from quadratic dispersion can lead to a large change in the second sound dispersion. Figure~\ref{fig:w-k} (c) shows the dispersion excluding flexural phonons. 
Comparing Fig.~\ref{fig:w-k} (b) and (c), we find that inclusion of flexural phonons drastically reduces the velocity of the drifting mode. Notably, in the two seminal papers on second sound in graphene, Ref.~\onlinecite{lee_hydrodynamic_2015} considered drifting mode, while Ref.~\onlinecite{cepellotti_phonon_2015} considered the driftless mode.

\begin{figure}
	\centering 
    \includegraphics[scale=0.45]{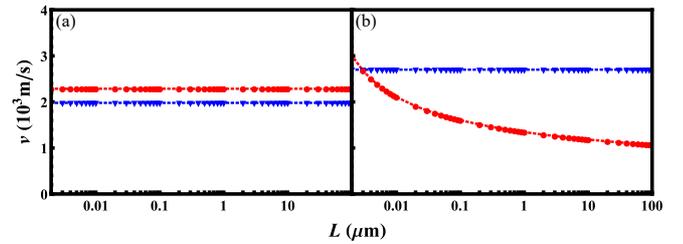}
	\caption{Size ($L$) dependence of drifting (red, circle) and driftless (blue, triangle) second sound velocities before (a) and after (b) applying the rotational symmetry.
	} 
	\label{fig:v-ss} 
\end{figure}

Figure \ref{fig:v-ss} shows the length dependence of the drifting (red) and driftless (blue) second sound velocity. Before the symmetrization, both modes show size-independent group velocity. The reduction of drifting mode velocity due to increasing $\rho^{(0)}_N \propto {\rm ln}L$ is observed for ideal quadratic phonon dispersion. This indicates that the logarithmic divergence of $\rho_N^{(0)}$ or length dependence of $v_{p}$ is easily destroyed by small deviation from ideal quadratic dispersion in the numerical calculations. This may explain the reason why convergent results can be obtained in previous numerical results\cite{lee_hydrodynamic_2015,lee_hydrodynamic_2017}. 

\begin{figure}
  \centering 
  \includegraphics[scale=0.4]{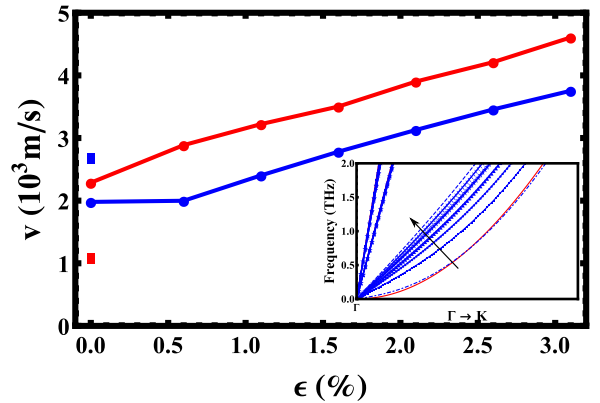} 
  \caption{Dependence of drifting (red) and driftless (blue) second sound velocity $v$ on the applied tensile strain $\epsilon = (a-a_0)/a_0$ for $L=50~\mu$m. The rectangular dots at $\epsilon = 0$ are results after applying rotational symmetry. In the inset, the red solid line shows the dispersion of flexural mode after symmetrization. The blue lines show how the dispersion changes with increasing tensile strain from 0\% to 3.1\% in the arrow direction. Hardening of the flexural mode with tensile strain is observed, while the transverse and longitudinal acoustic modes are less affected.}  
  \label{fig:vepsilon} 
\end{figure}

Since tensile strain can introduce a linear part to the dispersion to flexural phonons, we have plotted how it changes the velocity of drifting and driftless second sound in Fig.~\ref{fig:vepsilon}. Both of them increase with applied strain. This can be attributed to the increasing group velocity of flexural phonons with strain (inset). It is a special feature of 2D materials with flexural phonons and is, in principle, observable in experiments.  

Finally, we discuss implication of these results on the thermal conductivity of 2D materials. The presence of drift motion in Eq.~(\ref{eq:qm0}) generally results in larger thermal conductivity than that obtained from relaxation time approximation\cite{callaway_model_1959}. This has been used as one signature of hydrodynamic transport\cite{zhang_hydrodynamic_2020,cepellotti_phonon_2015}. For sufficiently weak damping, the collective drift motion sustains and the thermal conductivity may diverge. Our analysis here suggests that ideal quadratic dispersion of flexural phonons suppresses the drift motion in the long wave length limit. 
On the other hand, it has been predicted that thermal conductivity of 2D materials diverges logarithmically with system size. This has been supported both by numerical calculations\cite{Pereira13,bonini_acoustic_2012,lindsay_phonon_2014,fugallo_thermal_2014,kuang_unusual_2015,kuang_thermal_2016,gu_first-principles_2015} and experimental measurement\cite{Xu2014}. Numerical simulations further suggest that, divergent thermal conductivity emerges once there is small linear contribution to the flexural phonon dispersion. This seems inline with our prediction that drifting second sound emerges in the thermodynamic limit under the same condition. However, it has been shown\cite{bonini_acoustic_2012,gu_first-principles_2015} that thermal conductivity diverges even when the Boltzmann equation is solved under relaxation time approximation in which case hydrodynamic transport is absent. Thus, we argue that hydrodynamic phonon transport is not a necessary condition for divergent thermal conductivity. 
%Further numerical study is needed to clarify their connection. 

\section{Conclusions}
We have clarified the role of flexural phonons in 2D materials on the propagation of drifting and driftless second sound by developing a theory that takes both into account under equal footing. In addition to providing enlarged $N$-scattering phase space, an infinite effective inertial effect of flexural phonons, logarithmically divergent with system size, destroys the propagation of drifting second sound in the thermodynamic limit. On the other hand, the driftless second sound is less affected. We suggest that propagation of high frequency driftless second sound is possible even in diffusive system and may have been experimentally observed\cite{beardo_observation_2021}. This greatly extends the scope of materials where wave-like heat transport can be explored.

%On the other hand, for sufficiently weak damping of the drifting second sound, the thermal conductivity will diverge. Further quantitative study is needed for the connection, if any, between drift phonon motion and divergent thermal conductivity of 2D materials.

%But this suppression is fragile to small deviation from the quadratic form. Further quantitative study is needed for the connection if any between drift phonon motion and divergent thermal conductivity of 2D materials. %Our study of heat conduction from a hot spot seems supporting this connection\cite{zhang2020violation}.

\acknowledgements
 We thank Wu Li for assistance in recovering ideal quadratic dispersion of flexural phonons. We acknowledge funding support from National Natural Science Foundation of China (Grant No. 21873033), the National Key Research and Development Program of China (Grant No. 2017YFA0403501), the program for HUST academic frontier youth team. The computing resources are provided by Shanghai supercomputer center.

\appendix
%\onecolumngrid
%+++++++++++++++++++++++++++
%++++++++++++++++++++++++++++++++++++++++++++++
\setcounter{figure}{0}
\renewcommand{\thefigure}{S\arabic{figure}}
\renewcommand{\thetable}{S\arabic{table}}
%++++++++++++++++++++++++++++++++++++++++++++++

\section{\label{app:theory} Derivation of the hydrodynamic equations}

\subsection{Kinetic equation and conservation laws}
%\section{\label{sec:thermo}Thermodynamics}
%\section{Unified description of quasi-particle hydrodynamics in solids}
We consider a generalized Debye model to include the flexural phonons with quadratic dispersion $\omega_k= a k^2$, in additional to the linear mode $\omega_{k} = v_{g} k$. We define an effective mass 
$m^* = \hbar/(2a)$, so that
the energy of flexural phonons can be written as
$\hbar\omega_k = {\hbar^2 k^2}/(2m^*)$.
This form resembles that of non-relativistic particles. Correspondingly, phonons with linear dispersion could be thought as relativistic quasi-particles. 

To describe phonon transport in the system, we start from the Peierls-Boltzmann equation in the Callaway approximation [Eq.~(\ref{eq:callaway})].
%\cite{callaway_model_1959,allen_improved_2013}:
%\begin{equation}
%\frac{\partial{f}_{i\VK}}{\partial{t}}+\Vv_{i\VK}\cdot\nabla{f}_{i\VK}
%=-\frac{f_{i\VK}-f_{R,i\VK}^{}}{\tau_{R,i\VK}}-\frac{f_{i\VK}-f_{N,i\VK}^{}}{\tau_{N,i\VK}}.
%\label{eq:callaway}
%\end{equation}
%Here 
%\begin{equation}
%f_{R,i\VK}^{eq}=\frac{1}{{\rm exp}(\beta\hbar\omega_{i\VK})-1}
%\end{equation}
%is the equilibrium Bose-Einstein distribution, and
%\begin{equation}
% f_{N,i\VK}^{eq}=\frac{1}{{\rm exp}[\beta(\hbar\omega_{i\VK}-\hbar\VK\cdot\Vu)]-1}
%\end{equation}
%is the displaced distribution, with $\hbar$ the reduced Planck constant, $\beta=(k_BT)^{-1}$ the inverse temperature, $k_B$ the Boltzmann constant, $T$ the absolute temperature. 
%Here, $\tau_{R,i\VK}$ and $\tau_{N,i\VK}$ are the mode-resolved, $\VK$-dependent relaxation time due to $R$ and $N$ processes, respectively. $f_{i\VK}$ is the nonequilibrium phonon distribution, and $\Vv_{i\VK}$ is the phonon group velocity. The phonon index $i=L, N$, representing acoustic phonons with linear and quadratic dispersion, respectively. In the following, we will omit the wave vector $\VK$ and phonon branch index $i$ for brevity when there is no ambiguity. It is known that $R$-process takes the system to $\freq$, while in the presence of only $N$-process, the system relaxes to $\fneq$ with a characteristic velocity $\Vu$. 
Following the standard approach, we get the respective balance equations for energy and momentum density given in Eqs.~(\ref{eq:econ}-\ref{eq:pcon}).
%\begin{subequations}
%\begin{align}
%\frac{\partial{E}}{\partial{t}}+\nabla\cdot\Vq&=0, \label{eq:econ0}\\
%\frac{\partial{\VP}}{\partial{t}}+\nabla\cdot\Vphi&=-\AV{\tau^{-1}_{R}}_p\VP.
%\label{eq:pcon1}
%\end{align}
%\end{subequations}  
The 2D energy ($E$) and momentum ($\VP$) density are defined as
\begin{subequations}
\begin{align}
    E &=L^{-2}\sum_{i\VK} \hbar\omega_{i\VK} f_{i\VK}= \sum_i E_{i},\\
    \VP &= L^{-2}\sum_{i\VK} \hbar\VK f_{i\VK}= \sum_{i} \VP_i, 
\end{align}
\end{subequations}
while $\Vq$ and $\Vphi$ are their flux densities
\begin{align}
    \Vq &= L^{-2}\sum_{i\VK} \hbar\omega_{i\VK} \Vv_{i\VK} f_{i\VK}= \sum_{i\VK} \Vq_{i\VK} ,\\
    \Vphi &= L^{-2}\sum_{i\VK} \hbar\VK \Vv_{i\VK} f_{i\VK}= \sum_{i\VK} \Vphi_{i\VK}  .
\end{align}
Here, $L^2$ is the area of the sample.
We have introduced an averaged relaxation time defined as
\begin{subequations}
\begin{align}
    \label{eq:taurp}
    \AV{\tau^{-1}_{R}}_p &= \frac{\sum_{i\VK}\tau^{-1}_{R,i\VK}\hbar\VK(f_{i\VK}-f_{R,i\VK})}{\sum_{i\VK}\hbar
    \VK(f_{i\VK}-f_{R,i\VK})},\\
    \label{eq:taunp}
    \AV{\tau^{-1}_{N}}_p &= \frac{\sum_{i\VK}\tau^{-1}_{N,i\VK}\hbar\VK(f_{i\VK}-f_{N,i\VK})}{\sum_{i\VK}\hbar
    \VK(f_{i\VK}-f_{N,i\VK})}.
\end{align}
\end{subequations}
%Other types of average $\AV{\tau_R^{-1}}_q$, $\AV{\tau_N^{-1}}_q$,$\AV{\tau_R^{-1}}_\Phi$,$\AV{\tau_R^{-1}}_\Phi$ can be defined similarly. 
We have $\AV{\tau_N^{-1}}_p=0$ due to crystal-momentum conservation.  These averages are different, which is a manifestation of the fact that the same scattering process may relax physical quantities with different efficiency.
%As mentioned in Sec.~\ref{sec:intro}, we have ignored the scattering process that transfers energy from phonons to other degrees of freedom, i.e., electrons.  
%Although total phonon number is not conserved, in the following it is convenient to express other macroscopic quantities using it
%\begin{align}
%    n = L^{-2}\sum_{i\VK}f_{i\VK}= L^{-2}\sum_i n_i .
%\end{align}
%and its flux
%\begin{align}
%    \Vj = \sum_i \Vj_i = \sum_{i\VK} \Vv_{i\VK} f_{i\VK}.
%\end{align}

%------------------------------------
\subsection{Zeroth order approximation and Euler equation}
%------------------------------------
We follow the standard approach in deriving the hydrodynamic equations\cite{LandauBook6,guo_phonon_2015,shang_heat_2020}.
In the zeroth order approximation, the distribution function is approximated by 
\begin{align}
    f\approx f^{(0)} = \fneq.
    \label{eq:sbe}
\end{align}
Here, the superscript $(0)$ means zeroth order. This applies when $N$-scattering process is much stronger than $R$-scattering.
In the following, to derive the zeroth order physical quantities, we use the full form of $\fneq$ instead of the commonly used expansion 
%\begin{align}
    $\fneq \approx \freq + \beta \freq(\freq+1)\hbar\VK\cdot \Vu$.
    %\label{eq:smallu}
%\end{align}
The small-$u$ limit can be taken afterwards based on the full results. This is the main difference between present work and most of previous studies. We will show that this is convenient to understand the divergence encountered when dealing with flexural phonons.
In order to get closed expressions, we make the following approximation for the summation over $\VK$:
\begin{align}
    L^{-2}\sum_{\VK}f_{\VK}(\cdot) \approx \frac{1}{4\pi^2}\int_0^{2\pi}d\theta \int_0^{\infty}d\omega D(\omega)f_{\VK} (\cdot). 
    \label{eq:inta}
\end{align}
That is, we consider isotropic material in the temperature range $T\ll T_D$, with $T_D$ the Debye temperature. 
In the following we derive results for the number, energy, momentum and their corresponding fluxes in the zeroth order, for the linear and quadratic modes separately. Based on this, a generalized Euler equation is obtained.
%From this, all relevant macroscopic quantities, including $E$ and $\VP$, can be calculated. Substituting them into Eqs.~(\ref{eq:econ}-\ref{eq:pcon}), the zeroth order hydrodynamic equations can be obtained. 

\subsubsection{Number density}% and flux}
We start from the number density,
which is obtained by summing over all the allowed wave vectors. 
The mode dispersion matters in the summation. Straightforward calculation then yields
%\begin{subequations}
%\begin{align}
%    n_{L}^{(0)}& =  \frac{2}{(4\pi)^{d/2}}\frac{\Gamma(d)}{\Gamma(d/2)}{\rm Li}_{d}(\lambda) (1-\tu^2)^{-(d+1)/2} (\hbar v_g \beta)^{-d},\\
%    n_{N}^{(0)} &= \frac{}{(4\pi)^{d/2}}{\rm Li}_{d/2}(\lambda') (\hbar\beta a)^{-d/2}.
%\end{align}
%\end{subequations}
\begin{subequations}
\begin{align}
    n_{L}^{(0)}& =  \frac{\pi}{12}(1-\tu^2)^{-3/2} (\hbar v_g \beta)^{-2},\label{eq:nl0}\\
    n_{N}^{(0)} &= -\frac{1}{4\pi}{\rm ln}(1-\lambda') (\hbar\beta a)^{-1}.\label{eq:nn0}
\end{align}
\end{subequations}
Here, ${\bf \tu}=\Vu/v_g$ is the reduced velocity, $\lambda' = {\rm exp}(\beta \mu')$ with $\mu'=m^* u^2/2$  the effective chemical potential due to the drift motion.

We find that, in both cases, non-zero $\Vu$ gives rises to correction of the quasi-particle density, which is even order in $u$. For linear mode, we get an extra factor $(1-\tu^2)^{-3/2}$, which resembles that of the relativistic particles, with the velocity of light replaced by the group velocity of phonon quasi-particle. For the quadratic mode, the correction comes from the effective change of the chemical potential $\mu'$  due to the drift motion. 

Actually, this form of $n_N^{(0)}$ should be taken with caution. In the limit of $u\to 0$, $n_N^{(0)}$ diverges logarithmically. The reason behind this result is simple. The quadratic flexural phonons have a constant density of states in the long wave length limit. A singularity appears at $k$ where the Bose-Einstein distribution diverges, leading to divergent $n_N^{(0)}$. One more difficulty is that, when $\mu' > 0$, phonons with $\hbar\omega_k < \mu'$ has a unphysical negative population, meaning that these phonon modes can not be properly taken into account within the present kinetic theory. 
%In fact, all the singular behaviour of flexural phonons can be traced back to this. 
This kind of difficulty does not appear for linear phonon modes. We will discuss the consequence of divergent $n_N^{(0)}$. Meanwhile, we still use this formal result of Eq.~(\ref{eq:nn0}) for the expressions of other quantities. 

The above results of phonon number density shows the important difference between phonon quasi-particles and real particles with number conservation. Since phonons simply represent thermal excitation of the system, their number is not conserved.

%The quasi-particle flux can be obtained similarly. It takes a unified form  and does not depend on dispersion 
%\begin{align}
%    \Vj_i = \sum_{\VK} \Vv_{i\VK} f_{i\VK} = n_i \Vu.
%\end{align}

\subsubsection{Energy density and flux}
Following similar procedure, the zeroth order energy density can be obtained
%------------------
%\begin{subequations}
%\begin{align}
%E^{(0)}_{L}^{}&=A_L (d+\tilde{u}^2)(1-\tilde{u}^2)^{-(d+3)/2} (\hbar v_g \beta)^{-d} k_BT,\label{eq:el0}\\
%E^{(0)}_{N}^{} &=E_{N,U}^{}+E_{N,K}^{}=\frac{\Gamma(1+d/2)}{\Gamma(d/2)} n_{N}^{(0)} k_BT+\frac{1}{2}n_{N}^{(0)} m^* u^2.
%\label{eq:en0}
%\end{align}
%\end{subequations}
\begin{subequations}
\begin{align}
E^{(0)}_{L}&=\frac{1}{2\pi}{\rm Li}_3(\lambda) (2+\tilde{u}^2)(1-\tilde{u}^2)^{-\frac{5}{2}} (\hbar v_g )^{-2} (k_BT)^3 ,\label{eq:el0}\\
E^{(0)}_{N} &=E^{(0)}_{N,U}+E^{(0)}_{N,K}\nonumber\\
&=\frac{1}{4\pi}{\rm Li}_2(\lambda') (\hbar a)^{-1} (k_BT)^2+\frac{1}{2}\rho_N^{(0)} u^2, \label{eq:en0}
\end{align}
\end{subequations}
with $\rho^{(0)}_N= m^*n_{N}^{(0)}$, $\zeta(x)$ the Riemann Zeta function.
We have split $E_N$ into two terms $E^{(0)}_{N,U}$ and $E^{(0)}_{N,K}$. Their physical meaning is clear. The first term $E^{(0)}_{N,U}$ has the same form as the energy density in thermal equilibrium, except now $\lambda'> 1$ has an implicit $u$ dependence. The second term $E^{(0)}_{N,K}$ is a drift correction due to the collective motion of all the phonons with the  same velocity $\Vu$. Although $n_N^{(0)}$ itself diverges logarithmically when $u \to 0$, $E^{(0)}_{N,K}$ instead goes to zero due to the presence of an extra $u^2$. %Notably, $E^{(0)}_N$ is formally similar to that of classical non-relativistic gas particles with a common drift velocity $\Vu$. 

%Equations (\ref{eq:el0}-\ref{eq:en0}) show the different form of energy correction to the linear and quadratic modes due to collective drift of the phonon gas. For quadratic mode, the correction is similar to that of non-relativistic gases. But they are not exactly the same, since the phonon number is not conserved. The finite drift also modifies the number density $n_N^{(0)}$, resulting in its $u$ dependence. Correspondingly, for linear mode, the correction is similar to , but no exactly the same as, that of relativistic gases. This difference between phonon quasi-particles and the real particles lies in the fact that, phonons live in solids, which is always at rest in the laboratory frame. Additionally, the form of phonon-phonon interaction is always non-relativistic independence of the dispersion relation.

%In Appendix \ref{sec:appthermo} we give a brief account of the statistical thermodynamics of phonons. 
At this point, it is convenient to introduce 
%
%the enthalpy density of phonons, defined as
%\begin{align}
%    W = E + P, 
%\end{align}
%with $P$ the thermodynamic pressure.
%From the phonon equation of state in the zeroth order approximation (Appendix \ref{sec:appthermo}) and the expressions for the energy density, we obtain 
\begin{subequations}
\begin{align}
W^{(0)}_L &\equiv  3(1-\tu^2)^{-1}P_L^{(0)},\\
W^{(0)}_N &\equiv 2P^{(0)}_N+\frac{1}{2} \rho^{(0)}_N u^2,
\end{align}
\label{eq:defw}
\end{subequations}
with
\begin{subequations}
\begin{align}
P_L^{(0)} &= \frac{1}{2\pi}{\rm Li}_3(\lambda)(\hbar v_g)^{-2}(1-\tu^2)^{-3/2}(k_BT)^3,\\ 
P_N^{(0)} &= \frac{1}{4\pi} {\rm Li}_{2}(\lambda')(\hbar a)^{-1}(k_BT)^2.\label{eq:p0n}
\end{align}
\end{subequations}
$W^{(0)}$ and $P^{(0)}$ can be thought as the enthalpy density and pressure of the phonon gas, evaluated from the drifted Bose-Einstein distribution function [Eq.~(\ref{eq:sbe})]. From this, we can write the energy density in a different form
\begin{subequations}
\begin{align}
    E_L^{(0)} &= \frac{2+\tilde{u}^2}{1-\tilde{u}^2} P_L^{(0)}, \label{eq:el02}\\
    E_N^{(0)} &= P_N^{(0)} + \frac{1}{2}\rho_N^{(0)}u^2 .
\end{align}
\end{subequations}
The phonon energy flux is then written in a compact form
\begin{align}
\Vq_i^{(0)} = W_i^{(0)} \Vu,
\label{eq:q0}
\end{align}
which is applicable to both linear and quadratic modes.
%We note that the definition of $W^{(0)}$ resembles the enthalpy density of photon gas  $W^{(0)} = E^{(0)} + P^{(0)}$,  with $P^{(0)}$ given in Appendix~\ref{sec:appthermo}. 

\subsubsection{Momentum density and flux}
The momentum density of linear and quadratic modes takes quite different forms 
\begin{subequations}
\begin{align}
    \VP^{}_L &= v_g^{-2} \Vq_L^{}  \label{eq:pn0},\\
    \VP^{(0)}_N &=\rho_N^{(0)}\Vu.  \label{eq:pl0}
\end{align}
\end{subequations}
The linear dispersion leads to a general relationship between $\VP_L$ and $\Vq_L$, Eq.~(\ref{eq:pn0}), which is obtained from their definition and holds to all (quasi-)particles with linear dispersion. In the derivation of the Guyer-Krumhansl equation within the Debye model, this relationship plays a key role in relating the momentum density with the heat flux \cite{guo_phonon_2015}.
However, there is no such relationship between $\VP_N$ and $\Vq_N$ \cite{shang_heat_2020}. Instead, Eq.~(\ref{eq:pl0}) takes a same form as the traditional non-relativistic gas. It shows that the momentum density is simply the drift momentum $m^* \Vu$ times the number density $n_N^{(0)}$. Similar to $E_N^{(0)}$, $\VP_N^{(0)} \to 0$ when $\Vu \to 0$. 
If we define an effective mass density for the linear mode
\begin{align}
    \rho_L^{(0)} \equiv v_g^{-2} W_L^{(0)}, \label{eq:rhol}
\end{align}
$\VP_L^{(0)}$ and $\VP_N^{(0)}$ can be written in the same form.
Consequently, the momentum flux is written in a unified form 
\begin{align}
    \Phi_{i,mn}^{(0)} &= P_i^{(0)}\delta_{mn} + \rho_i^{(0)} u_m u_n.
    \label{eq:phi0}
\end{align}

\subsubsection{The generalized Euler equation}
Substituting the momentum flux Eq.~(\ref{eq:phi0}) into its balance equation,  we arrive at  
\begin{align}
%(\partial_t + \Vu\cdot \nabla) \rho^{(0)}\Vu + (\nabla\cdot \Vu)\rho^{(0)}\Vu +  \nabla P^{(0)} &= -\AV{\tau_{R}^{-1}}_p \rho^{(0)}\Vu.\\
(\partial_t + \Vu\cdot \nabla) \Vu + (\nabla\cdot \Vu)\Vu +  \nabla P^{(0)}/\rho^{(0)} &= -\AV{\tau_{R}^{-1}}_p\Vu.
\label{eq:euler}
\end{align}
%\begin{subequations}
%\begin{align}
%\label{eq:eulerl}
%v_g^{-2}{W^{(0)}} (\partial_t + \Vu\cdot \nabla) \Vu + v_g^{-2}[\nabla\cdot ({W^{(0)}}\Vu)]\Vu + \nabla P &= -v_g^{-2}{\tau_{R,p}}^{-1} {W^{(0)}} \Vu,\\
%nm^*(\partial_t + \Vu\cdot \nabla) \Vu + [\nabla\cdot (nm^*\Vu)]\Vu +  \nabla P &= -{\tau_{R,p}}^{-1} nm^* \Vu.
%\label{eq:eulern}
%\end{align}
%\end{subequations}
%We have kept the term representing momentum-destroying scattering 
%\begin{align}
%    \tau_{R,p}^{-1} =(\tau_{R,p,L}^{-1}\rho_L^{(0)}+\tau_{R,p,N}^{-1}\rho_N^{(0)})/\rho^{(0)},
%\end{align}
%with $\rho^{(0)}=\rho^{(0)}_L+\rho^{(0)}_N$.
Equation~(\ref{eq:euler}) is the generalized Euler equation for phonons, where the driving force is $\nabla P^{(0)}$. Here $\rho^{(0)}$ without sub-index represents the total `mass' density, similarly for other physical quantities. In the analysis of second sound, we will keep only the linear-in-$u$ terms in the equations. 
Similarly, substituting Eq.~(\ref{eq:q0}) into Eq.~(\ref{eq:econ}), we obtain the energy balance equation for both types of modes
\begin{align}
\frac{\partial{E^{(0)}}}{\partial{t}}+\nabla\cdot (W^{(0)}\Vu)&=0. \label{eq:econ2}
\end{align}

\subsubsection{Comparison to standard relativistic and non-relativistic hydrodynamics}
\label{subsec:renonrel}
In the above subsections, we have  written our results for the linear and quadratic phonons in similar forms as those of the standard relativistic and non-relativistic hydrodynamic equations\cite{LandauBook6}, respectively. Especially, the energy, momentum density and their corresponding fluxes [Eqs.~(\ref{eq:el02}-\ref{eq:phi0})] take the standard form. However, they do have one important difference. Here, all quantities with superscript $(0)$ represent results obtained from the drifted Bose-Einstein distribution function [Eq.~(\ref{eq:sbe})], instead of the equilibrium one as in standard hydrodynamics. Thus, strictly speaking, the linear and quadratic phonons do not fulfill the Lorentz or Galilean invariance. This has been discussed in the context of relativistic hydrodynamics of electrons in graphene\cite{narozhny_electronic_2019}.
However, when considering second sound propagation, we will keep only the linear-in-$u$ terms. In that case, the quantities with superscript $(0)$ are obtained from the equilibrium Bose-Einstein distribution, and the Lorentz and Galilean invariance are recovered for the linear and quadratic phonons, respectively. 

\subsection{Transport coefficients and dissipation}
The obtained generalized Euler equation does not include any internal dissipation due to $N$-scattering. They are included in the higher order corrections. In Sec.~\ref{app:transport} we give detailed derivation based on kinetic theory up to the first order in the small parameter 
\begin{align}
    \varepsilon_{i\VK}=\tau_{N,i\VK}/\tau_{R,i\VK}.
\end{align}
Here, we follow the general phenomenology, and write the momentum flux $\Phi_{mn}^{(1)}$ into the following form
\begin{align}
\label{eq:phi1g}
    \Phi_{mn}^{(1)} = -\eta \left(\partial_{x_n}u_m+\partial_{x_m}u_n-\frac{2}{D}\delta_{mn}\partial_{x_i}u_i\right) - \xi \delta_{mn} \partial_{x_i}u_i,
\end{align}
with $D$ the system dimension ($D=2$ here), $\eta$ and $\xi$ the shear and bulk viscosity, respectively. The corresponding momentum balance equation including dissipation takes the general form:
%\begin{align}
%(\partial_t + \Vu\cdot \nabla) \rho^{(0)}\Vu + (\nabla\cdot \Vu)\rho^{(0)} \Vu +  \nabla P &= \eta \nabla^2 \Vu -\left(\eta-\xi+\frac{2}{d}\eta\right)\nabla(\nabla\cdot\Vu)- \AV{\tau^{-1}_{R}}_p\rho^{(0)}\Vu.
%\label{eq:ns}
%\end{align}
\begin{align}
%&(\partial_t + \Vu\cdot \nabla) \rho^{(0)}\Vu + (\nabla\cdot \Vu)\rho^{(0)} \Vu +  \nabla P = \eta \nabla^2 \Vu \nonumber\\
%&+\left[\left(1-\frac{2}{D}\right)\eta+\xi\right]\nabla(\nabla\cdot\Vu)- \AV{\tau^{-1}_{R}}_p\rho^{(0)}\Vu.\\
&(\partial_t + \Vu\cdot \nabla) \Vu + (\nabla\cdot \Vu)\Vu +  \nabla P/\rho^{(0)}  = \eta /\rho^{(0)} \nabla^2 \Vu \nonumber\\
&+\left[\left(1-\frac{2}{D}\right)\eta+\xi\right] /\rho^{(0)} \nabla(\nabla\cdot\Vu)- \AV{\tau^{-1}_{R}}_p\Vu.
\label{eq:ns}
\end{align}
We will derive the viscosity coefficients $\eta$ and $\xi$ [Eqs.~(\ref{eq:eta}-\ref{eq:xi})] in 
App.~\ref{app:transport}.
\subsection{Energy flux equation and thermal conductivity}
The energy flux balance equation can be obtained by multiplying $\hbar\omega_{ik} \Vv_{ik} \tau_{c,i\VK}$ to Eq.~(\ref{eq:callaway}) and performing the summation over $i\VK$
\begin{align}
    &\partial_t \sum_{i\VK} (1-\chi_{i\VK}) \tau_{R,i\VK} \Vq_{i\VK} + \Vq =  \sum_{i\VK}\chi_{i\VK}\Vq_{i\VK}^{(0)}\nonumber\\
    &-\sum_{i\VK} (1-\chi_{i\VK}) \times \partial_T f_{i\VK}\tau_{R,i\VK} \hbar\omega_{i\VK} \Vv_{i\VK} \Vv_{i\VK}  \cdot \nabla T.  
    \label{eq:q00-1}
\end{align}
The combined relaxation time is defined as 
\begin{align}
\tau_{c,i\VK}^{-1} = \tau_{N,i\VK}^{-1}+\tau_{R,i\VK}^{-1}.
\end{align}
It shows that the total energy current includes two contributions: one due to the collective phonon drift motion, the other due to the temperature gradient. Their relative contributions are weighted by two factors $\chi_{i\VK}=(1+\varepsilon_{i\VK})^{-1}$ and $1-\chi_{i\VK}$, respectively. 
This equation is valid in the full range of $\varepsilon_{i\VK}$ and can be expanded over $\varepsilon_{i\VK}$ in the hydrodynamic regime when $\varepsilon_{i\VK} \ll 1$. The zeroth order result gives the null result $\Vq^{(0)} = \Vq^{(0)}$, while the first order equation is exactly Eq.~(\ref{eq:q1}). 
To proceed, we write Eq.~(\ref{eq:q00-1}) into a compact form 
\begin{align}
    \AV{\tau_{c}}_q \partial_t \Vq + \Vq = \chi W^{(0)}\Vu - \Vkap \cdot \nabla T. 
    \label{eq:qm}
\end{align}
We have introduced an averaged characteristic relaxation time of $\Vq$ 
\begin{align}
    \AV{\tau_c}_q = \frac{\sum_{i\VK}\frac{\varepsilon_{i\VK}}{1+\varepsilon_{i\VK}}\tau_{R,i\VK}\Vq_{i\VK}}{\sum_{i\VK}\Vq_{i\VK}}\approx \frac{\sum_{i\VK}\frac{\varepsilon_{i\VK}}{1+\varepsilon_{i\VK}}\tau_{R,i\VK}\Vq^{(0)}_{i\VK}}{\sum_{i\VK}\Vq^{(0)}_{i\VK}},
    \label{eq:qtau}
\end{align}
the averaged quantities 
\begin{align}
    \label{eq:eq0}
    \chi = \frac{\sum_{i\VK}\Vq_{i\VK}^{(0)} (1+\varepsilon_{i\VK})^{-1}}{\sum_{i\VK}\Vq_{i\VK}^{(0)}},
\end{align}
thermal conductivity from relaxation time approximation 
\begin{align}
\label{eq:kappa}
\kappa&=L^{-2}\sum_{i\VK} \hbar\omega_{i\VK} v_{i\VK} v_{i\VK} \tau_{c,i\VK} \partial_T f_{i\VK} \nonumber\\
&\equiv L^{-2}\sum_{i\VK} \hbar\omega_{i\VK} v_{i\VK} v_{i\VK} \AV{\tau_c}_{\kappa} \partial_T f_{i\VK},
\end{align}
 where the second equality defining an average relaxation time $\AV{\tau_c}_\kappa$. We note that, due to the $i\VK$ dependence of $\tau_{c,i\VK}$, the two averages $\AV{\tau_c}_\kappa$ and $\AV{\tau_c}_q$ are different.

\section{Derivation of the transport coefficients}
\label{app:transport}
To consider dissipation in the hydrodynamic equations, we need to include the first order correction. By performing an expansion over ${\varepsilon_{i\VK}}$, the  first order correction to distribution function is\cite{guo_phonon_2015}
\begin{align}
    f^{(1)} &= {\varepsilon} (\freq -\fneq) - {\tau_{N}}(\partial_t f_N + \Vv\cdot\nabla f_N)\nonumber\\
    &= \varepsilon (\freq-\fneq)-\tau_N[\partial_T \fneq (\partial_t T+\Vv\cdot \nabla T)\nonumber\\
    &-\partial_{\hbar\omega} \fneq (\hbar\VK\cdot \partial_t \Vu + \Vv\cdot(\hbar\VK\cdot \nabla) \Vu)].\label{eq:f1}
\end{align}
The corresponding first order corrections to the momentum and energy flux are
\begin{align}
    \Phi^{(1)}_{mn} &= \sum_{i\VK} \hbar k_m v_{i\VK,n} f^{(1)}_{i\VK}, \\
    q^{(1)}_n &= \sum_{i\VK} \hbar \omega_{i\VK} v_{i\VK,n} f^{(1)}_{i\VK}.
\end{align}
In principle, we can divide $f^{(1)}$ into even and odd (in $\VK$) contributions. They contribute to $\Vphi^{(1)}$ and $\Vq^{(1)}$, respectively. The full evaluation of $\Vphi^{(1)}$ and $\Vq^{(1)}$ using Eq.~(\ref{eq:f1}) is quite cumbersome. Here, as in traditional hydrodynamics, we only consider terms that are first order in $\Vu$ and the deviations $\delta u$, $\delta T$.  
%---%The $\VK$ dependence of $\varepsilon_\VK$ and $\tau_{N,\VK}$ make the analysis difficult. To proceed, we can introduce the following averaged quantities 
%---%\begin{align}
%---%    \AV{\varepsilon}_{q} &= \frac{\sum_{i\VK}\varepsilon_\VK (\freq-\fneq)\hbar\omega_{i\VK} \Vv_\VK}{\sum_{i\VK} (\freq-\fneq)\hbar\omega_{i\VK} \Vv_\VK},\\
%---%    \AV{\tau_N}_{q} &=\frac{\sum_{i\VK}\tau_{N,\VK}(\partial_t f_N + \Vv\cdot\nabla f_N)\hbar\omega_{i\VK}\Vv_{i\VK}}{\sum_{i\VK}(\partial_t f_N + \Vv\cdot\nabla f_N)\hbar\omega_{i\VK}\Vv_{i\VK}},
%---%\end{align}
%---%when considering $\Vq^{(1)}$. Similar quantities can be defined 
%---%\begin{align}
%---%    \AV{\varepsilon}_{\Phi} &= \frac{\sum_{i\VK}\varepsilon_\VK (\freq-\fneq)\hbar\VK \Vv_\VK}{\sum_{i\VK} (\freq-\fneq)\hbar\VK \Vv_\VK},\\
%---%    \AV{\tau_N}_{\Phi} &=\frac{\sum_{i\VK}\tau_{N,\VK}(\partial_t f_N + \Vv\cdot\nabla f_N)\hbar\VK\Vv_{i\VK}}{\sum_{i\VK}(\partial_t f_N + \Vv\cdot\nabla f_N)\hbar\VK\Vv_{i\VK}},
%---%\end{align}
%---%when considering $\Vphi^{(1)}$.
The distribution function can then be split into odd ($f^{(1)}_o$) and even ($f^{(1)}_e$) parts
\begin{align}
f^{(1)}&\approx f^{(1)}_{e}+f^{(1)}_{o},
\label{eq:f1-1}
\end{align}
where
\begin{align}
f^{(1)}_{e} =&\tau_{N,i\VK} 
\left[\partial_{\hbar\omega}\freq\hbar  \VK \cdot (\Vv\cdot \nabla)\Vu- \partial_{T}\freq\partial_t T\right],\label{eq:f1e}\\
%f^{(1)}_{o}&=  \varepsilon_{i\VK} \partial_{\hbar\omega}\freq \hbar\VK\cdot \Vu +\tau_{N,i\VK}\nonumber\\ &\times\left(\partial_{\hbar\omega_{}}\freq\hbar  \VK\cdot\partial_t \Vu -\partial_{\hbar\omega_{}}\freq  \Vv\cdot \nabla T\right).\label{eq:f1o}\\
f^{(1)}_{o}=&  \varepsilon_{i\VK} \partial_{\hbar\omega}\freq \hbar\VK\cdot \Vu +\tau_{N,i\VK}\partial_{\hbar\omega_{}}\freq\nonumber\\ &\times\left(\hbar  \VK\cdot\partial_t \Vu -  \Vv\cdot \nabla T\right).\label{eq:f1o}
\end{align}

\subsubsection{Momentum flux}
The momentum flux is obtained from $f^{(1)}_e$ as
%\begin{subequations}
\begin{align}
\Phi_{mn}^{(1)} 
= &-\frac{1}{2}\sum_{i\VK}\tau_{N,i\VK} \alpha_i W_{i\VK}^{(0)} \partial_{x_{i}}u_j\nonumber\\
&\times(\delta_{ij}\delta_{mn}+\delta_{im}\delta_{nj}+\delta_{in}\delta_{mj})\nonumber\\
&-\sum_{i\VK}\tau_{N,i\VK} \alpha_i C_{i\VK}^{(0)} \partial_{t}T \delta_{mn},\nonumber\\
\equiv &-\frac{1}{4}\AV{\tau_N}_{\Phi} (W_L^{(0)}+2W_N^{(0)}) \partial_{x_{i}}u_j \nonumber\\
&\times(\delta_{ij}\delta_{mn}+\delta_{im}\delta_{nj}+\delta_{in}\delta_{mj})\nonumber\\
&-\frac{1}{2}\AV{\tau_N}_{\Phi}\partial_T (E_L^{(0)}+2E_N^{(0)}) \partial_{t}T \delta_{mn},
\label{eq:phi1}
\end{align}
Here, 
%$\gamma_i=1$ for linear mode  and $\gamma_i=2$ for quadratic flexural mode,  and 
$C_{i\VK}^{(0)}$, $W_{i\VK}^{(0)}$ are the mode resolved heat capacity and enthalpy function, respectively. We have defined the average relaxation time $\AV{\tau_N}_\Phi$ through the second equality.
%\end{subequations}
%In arriving at these results, we have defined a characteristic relaxation time
%\begin{align}
%    \AV{\tau_{N}}_{\Phi} = \frac{\sum_{i\VK}\hbar \VK \Vv_\VK \partial_{\hbar\omega_k}\freq \left(\hbar  \VK \cdot [(\Vv\cdot \nabla)\Vu]+ \hbar\omega \partial_t T/T\right)\tau_{N,i\VK}}{\sum_{i\VK}\hbar \VK \Vv_\VK \partial_{\hbar\omega_k}\freq \left(\hbar  \VK \cdot [(\Vv\cdot \nabla)\Vu]+ \hbar\omega \partial_t T/T\right)}.
%\end{align}
%These results show that the momentum flux is generated from two sources. The first term corresponds to a velocity gradient, and the second term corresponds to a time dependent temperature field. 
Comparing with the general form of $\Phi_{mn}^{(1)}$ [Eq.~(\ref{eq:phi1g})], we obtain the expressions for the shear ($\eta$) and bulk ($\xi$) viscosity 
\begin{subequations}
\begin{align}
    \label{eq:eta}
    \eta&=\frac{1}{4}\langle\tau_{N}\rangle_{\Phi} (W_L^{(0)}+2W_N^{(0)}),\\
    \xi&= \frac{1}{2}\left(\frac{1}{D}-\frac{1}{2}\right)\AV{\tau_{N}}_{\Phi} (W_L^{(0)}+2W_N^{(0)}).
    \label{eq:xi}
\end{align}
\end{subequations}
%\revision{We get a zero bulk viscosity for 2D case?}

\subsubsection{Energy flux}
The energy flux is obtained similarly from the odd correction $f_o^{(1)}$
%\begin{subequations}
\begin{align}
\Vq^{(1)} 
&= -\sum_{i\VK}\varepsilon_{i\VK}W_{i\VK}^{(0)}\Vu -\sum_{i\VK}\tau_{N,i\VK}W_{i\VK}^{(0)} \partial_{t}\Vu \nonumber\\
&-\sum_{i\VK} \tau_{N,i\VK}\alpha_i\hbar\omega_{i\VK}\partial_T f_{R,i\VK} v^2_{i\VK}\nabla T . \label{eq:q1}
%\Vq_{L}^{(1)} 
%&= -\AV{\varepsilon}_{q} W^{(0)}_L\Vu -\AV{\tau_{N}}_{q} W_L^{(0)} \partial_{t}\Vu -\frac{1}{2}\AV{\tau_{N}}_{q} v_g^2 \partial_{T}E^{(0)}_L\nabla T \label{eq:q1l}\\
%\Vq_{N}^{(1)}
%&=-\AV{\varepsilon}_{q} W^{(0)}_N\Vu-\AV{\tau_{N}}_{q}W^{(0)}_N\partial_{t} \Vu-\AV{\tau_{N}}_{q}v_g^2 \partial_{T}E^{(0)}_L\nabla T \label{eq:q1n}
\end{align}
%\end{subequations}
This result can also be obtained by expansion of the full expression [Eq.~(\ref{eq:q00-1})] to the first order in $\varepsilon_{i\VK}$. It is understood that, the $u\to 0$ limit of the zeroth order quantities  $W^{(0)}$ and $E^{(0)}$ in Eqs.~(\ref{eq:phi1}-\ref{eq:q1}) should be used to be consistent with the approximation used in Eqs.~(\ref{eq:f1-1}-\ref{eq:f1o}).
%We have defined
%\begin{align}
%    \VK \varepsilon_ q = \frac{\sum_{i\VK}\varepsilon_\VK \hbar\omega_k \VK \VK \partial_{\hbar\omega_k}\freq}{\sum_{i\VK}\hbar\omega_k \VK\partial_{\hbar\omega_k}\freq},
%\end{align}
%and
%\begin{align}
%    \AV{\tau_{N}}_{q} = \frac{\sum_{i\VK}\tau_{N,i\VK}\hbar\omega_k \VK \left(\hbar  \VK\cdot\partial_t \Vu+ \hbar\omega \Vv\cdot \nabla T/T\right)\partial_{\hbar\omega_k}\freq }{\sum_{i\VK}\hbar\omega_k \VK \left(\hbar  \VK\cdot\partial_t \Vu+ \hbar\omega \Vv\cdot \nabla T/T\right)\partial_{\hbar\omega_k}\freq}.
%\end{align}

%The energy balance equation including the first order correction reads
%\begin{align}
%    \partial_t E^{(0)} + (1-\AV{\varepsilon}_{q})\nabla \cdot (W^{(0)} \Vu) -\AV{\tau_{N}}_{q} \nabla \cdot (W^{(0)}\partial_t \Vu)-\frac{1}{2}(g+2)\AV{\tau_{N}}_{q} v_g^2 C_L \nabla^2 T &= 0.\label{eq:ebl}
%\end{align} 
%Here, $g$ is the number of the linear modes and $C_L$ is the specific heat capacity of one linear mode. Actually, $v_g^2 C_L = 2\zeta(3)k_B^3 T^2/\pi\hbar^2$ does not depend on the dispersion relation of the phonon mode. Given the same relaxation time, the heat conductivity from flexural mode with quadratic dispersion is twice that of the linear mode. 

\section{Comparison between second sound in helium \Rmnum{2} and in solids}
\label{app:helium-ss}
Here, for completeness, we present a comparison between second sound of helium \Rmnum{2} and phonon system in solids. 
%\subsection{Second sound in Helium \Rmnum{2}}
%The phenomenological two-fluid model developed by Tisza and Landau explained the superfluid properties of helium \Rmnum{2}.
%It also predicted the existence of second sound, which was later experimentally confirmed by
%Peshkov\cite{RD2009}. 
The first and second sound in helium \Rmnum{2} can be understood from linearized, non-dissipative version of Landau’s macroscopic hydrodynamic equations\cite{landau1941two}:
\begin{align}
\frac{\partial \rho}{\partial t}+\nabla \cdot \bm{j}&=0,
\label{mass-con}\\
\frac{\partial \bm{j}}{\partial t}+\nabla P&=0,
\label{momentum-con}\\
\frac{\partial \rho S}{\partial t}+\rho S \nabla \cdot \bm{v}_{n}&=0,
\label{entropy-con}\\
\frac{\partial \bm{v}_{s}}{\partial t}+\nabla \mu&=0.
\label{potential-flow}
\end{align}
Here, the subscripts $n$ and $s$ represent normal and super ﬂuid, respectively.
Equation (\ref{mass-con}) is a result of mass conservation, with $\rho$ the total mass density and
$j$ the mass ﬂux or momentum density. Equation (\ref{momentum-con}) represents the conservation
of momentum, with $P$ the pressure. Equation (\ref{entropy-con}) means that the total entropy
($\rho S$) is conserved, since we have ignored the dissipative processes. One important
point is that the entropy is only carried by normal ﬂuid, thus $\bm{v}_n$. The fourth
equation is special to super ﬂuid. It represents the potential ﬂow ($\mu$ the chemical
potential) of super ﬂuid. Landau obtained this equation from the requirement
of $\nabla \times \bm{v}_s=0$ . We now know that $\bm{v}_s \propto \nabla \Phi$, where $\Phi$ is the phase of the Bose-Einstein condensed superﬂuid wavefunction.

From Eqs.~(\ref{mass-con}-\ref{momentum-con}), we can get the equations of motion for $\bm{v}_n$ and $\bm{v}_s$, respectively,
\begin{align}
\rho_{n} \frac{\partial \bm{v}_{n}}{\partial t}+\frac{\rho_{n}}{\rho} \nabla P+\rho_{s} S \nabla T &=0, \\
\rho_{s} \frac{\partial \bm{v}_{s}}{\partial t}+\frac{\rho_{s}}{\rho} \nabla P-\rho_{s} S \nabla T &=0.
\end{align}
We can also write them in terms of the relative velocity of the normal and super ﬂuid $\bm{w}=\bm{v}_n-\bm{v}_s$ and obtain
\begin{align}
\rho \frac{\partial S}{\partial t}+\rho_{s} S \nabla \cdot \bm{w} &=0, 
\label{relative-entropy}\\
\frac{\partial \bm{w}}{\partial t}+\frac{\rho}{\rho_{n}} S \nabla T &=0.
\label{relative-temperature}
\end{align}
We see that, the in-phase motion of the normal and superﬂuid couples to pressure (Eqs.~\ref{mass-con}-\ref{momentum-con}), while the out-of-phase/relative motion couples to entropy and temperature (Eqs.~\ref{relative-entropy}-\ref{relative-temperature}). As a result, we obtain
\begin{align}
\frac{\partial^{2} \rho}{\partial t^{2}}-\nabla^{2} P=0, \\
\frac{\partial^{2} S}{\partial t^{2}}-\frac{\rho_{s} S^{2}}{\rho_{n}} \nabla^{2} T=0.
\end{align}
We now show that the former group gives rise to the first sound, and the latter group gives the second sound. Writing $\rho=\rho(P,T)$ and $S=S(P,T)$ in terms of $P$ and $T$,
%\begin{align}
%d \rho &=\left(\frac{\partial \rho}{\partial T}\right)_{P} d T+\left(\frac{\partial \rho}{\partial P}\right)_{T} d P, \\
%d S &=\left(\frac{\partial S}{\partial T}\right)_{P} d T+\left(\frac{\partial S}{\partial P}\right)_{T} d P.
%\end{align}
%From this, 
%we get:
%\begin{align}
%\left(\frac{\partial \rho}{\partial T}\right)_{P} \frac{\partial^{2} T}{\partial t^{2}}+\left(\frac{\partial \rho}{\partial P}\right)_{T} \frac{\partial^{2} P}{\partial t^{2}}-\nabla^{2} P &=0 ,
%\label{rewrite-rho}\\
%\left(\frac{\partial S}{\partial T}\right)_{P} \frac{\partial^{2} T}{\partial t^{2}}+\left(\frac{\partial S}{\partial P}\right)_{T} \frac{\partial^{2} P}{\partial t^{2}}-\frac{\rho_{s} S^{2}}{\rho_{n}} \nabla^{2} T &=0.
%\label{rewrite-S}
%\end{align}
%
%We now suppose there is small perturbation of $T$ and $P$ around its equilibrium values as:
%\begin{equation*}
%T=T_{0}+\delta T, \quad P=P_{0}+\delta P.
%\end{equation*}
%with
%\begin{equation*}
%\delta T, \delta P \propto e^{-i \omega(t-x/v)}.
%\end{equation*}
%Substituting into Eqs.~(\ref{rewrite-rho}-\ref{rewrite-S}) we get:
%
and performing a linear analysis, we get 
\begin{align}
\left(\frac{\partial \rho}{\partial T}\right)_{P} \delta T+\left(\frac{1}{v_{1}^{2}}-\frac{1}{v^{2}}\right) \delta P&=0, \\
\frac{v_{2}^{2} C_{}}{T}\left(\frac{1}{v_{2}^{2}}-\frac{1}{v^{2}}\right) \delta T+\left(\frac{\partial S}{\partial P}\right)_{T} \delta P&=0,
\end{align}
with
\begin{align}
v_{1}=\sqrt{\frac{\partial P}{\partial \rho}} , \quad
v_{2}=\sqrt{\frac{T S^{2} \rho_{s}}{C \rho_{n}}}.
\end{align}
The condition that the above two equations have solutions gives
\begin{equation}
\left(1-\frac{v_{1}^{2}}{v^{2}}\right)\left(1-\frac{v_{2}^{2}}{v^{2}}\right)
%=\left(\frac{\partial \rho}{\partial S}\right)_{P}\left(\frac{\partial S}{\partial \rho}\right)_{T}
%=\left(\frac{\partial T}{\partial P}\right)_{S}\left(\frac{\partial P}{\partial T}\right)_{\rho}
=0.
%=\frac{C_{P}-C_{T}}{C_{P}}.
\end{equation}
%To arrive at this result, we have used the relation between $C_P$ and $C_V$
%\begin{equation*}
%C_{P}-C_{V}=T\left(\frac{\partial S}{\partial V}\right)_{T}\left(\frac{\partial V}{\partial T}\right)_{P}=T V \alpha_{P}^{2} / \kappa_{T},
%\end{equation*}
%and a series of Maxwell relations. 
%Here, $\alpha_P$ is the thermal expansion coefficient
%at constant pressure and $\kappa_T$ is the isothermal compressibility. 
Here, we have ignored the difference between heat capacity at constant volumen and that at constant pressure. We get two sound solutions, whose velocities are given by $v_1$ and $v_2$. We recognize $v_1$ as velocity of the normal first sound. $v_2$ is then that of the second sound. Laudau showed that if we consider only phonon contribution (ignoring roton), we have
\(v_{2}=v_{1} / \sqrt{3}\).
%From Eqs.~(\ref{rewrite-rho}-\ref{rewrite-S}), we see that the first sound couples to $P$ and the second sound couples to $T$ . %Even if we take $C_P$ and $C_V$ differently, the first sound still couples mainly to the pressure, similar to the normal sound, and the second sound couples mainly to temperature/entropy. 
%This is illustrated schematically in the figure below. 

The most important point to present this analysis is to show that, the existence of second sound in helium \Rmnum{2} relies on the relative out-of-phase motion of the normal and super ﬂuid ($w$). Thus, second sound in helium \Rmnum{2} is a result of quantum mechanical effect, since its appearance relies on the presence of Bose-Einstein condensation.

\begin{figure} 
  \centering 
  \includegraphics[scale=0.45]{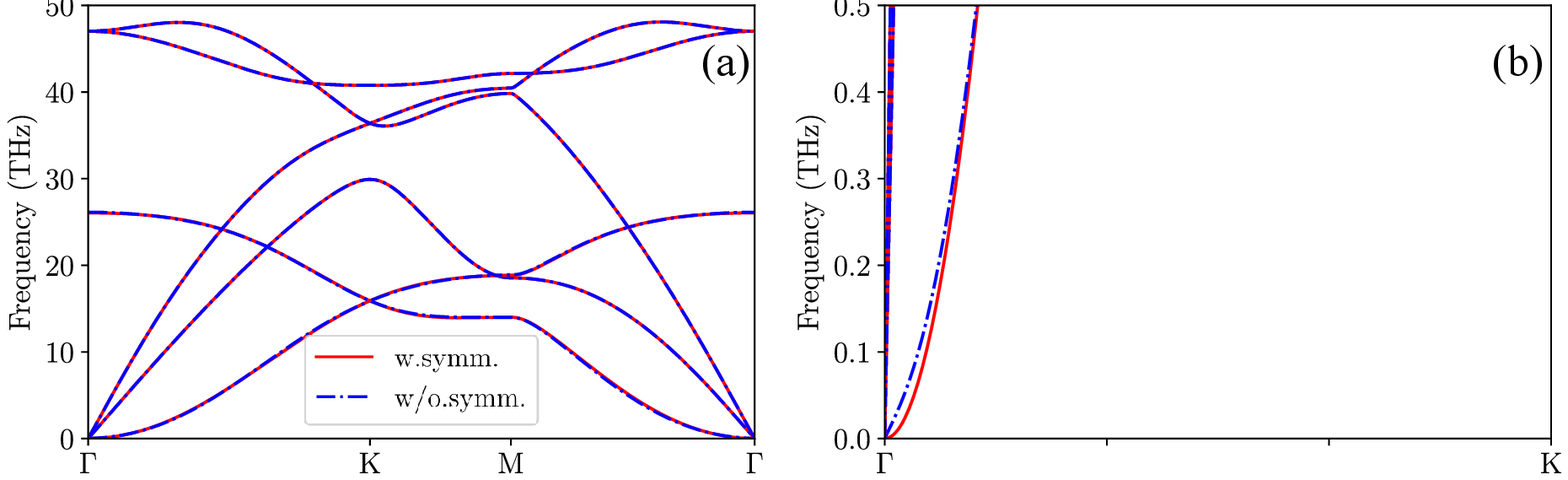} 
  \caption{(a) Phonon dispersion of graphene along high symmetry directions. (b) Zoom in of the dispersion near $\Gamma$ in (a). The blue and red lines correspond to results before and after the symmetrization of the force constants, which guarantees the translational and rotational symmetry are fulfilled. The ideal quadratic dispersion of ZA mode is recovered after the symmetrization.}
  \label{fig:wk-com} 
\end{figure}

%\subsection{Phonon second sound in solids}
%In solid state system, we can use the phonon gas model to study its thermal transport properties. Phonons are quasi-particles due to collective motion of atoms and are the most important thermal excitations in dielectrics. Different from real particles, they do not have number conservation.
%
%When crystal momentum conserving normal scattering is dominant over the momentum non-conserving scattering, phonons can relax to the drifted Bose-Einstein distribution
%$f_{N, i \bm{k}}=\left\{{\rm exp}\left([\beta_{}\hbar \omega_{i \VK}-\VK \cdot \Vu)\right]-1\right\}^{-1}$,
%with a common drift velocity $\Vu$. In the following, we show that this drift velocity takes a similar role as the relative velocity $\bm{w}$ in helium \Rmnum{2} and leads to the drifting second sound in phonon gas.
%
The similarity between second sound in helium \Rmnum{2} and in solids can be explained by considering the simple Debye model with one linear phonon branch $\omega=v_g k$. We assume that the phonon system follows the drifted distribution $f_{N,i\bm{k}}$ due to dominant normal scattering process.
In the linear-in-$\bm{u}$ limit, if we ignore all the dissipative processes, the energy ﬂux $\bm{q}$ and the momentum density $\bm{p}$ are proportional to $\Vu$, with
$\bm{q}=v_{g}^{2} \bm{p}=W^{(0)} \bm{u}$.
The energy balance equation then becomes:
\begin{equation}
\frac{\partial E^{(0)}}{\partial t}+W^{(0)} \nabla \cdot \bm{u}=0.
\label{energy-balance}
\end{equation}
We also have the momentum balance equation
\begin{equation}
\partial_{t} \bm{p}+\nabla P^{(0)}=0 \rightarrow  W^{(0)} v_{g}^{-2} \partial_{t} \bm{u}+\nabla P^{(0)}=0.
\label{momentum-balance}
\end{equation}
As discussed in Subsec.~\ref{subsec:renonrel}, although $W^{(0)}, E^{(0)}, P^{(0)}$ are calculated from the drifted distribution, their difference with the equilibrium value is at least second order in $u$. Thus, here to the lowest order, they can be taken as the equilibrium value. For $D$-dimensional phonon gas with linear dispersion, we have the following relations:
\begin{align}
E^{(0)}&=D P^{(0)},\\ 
W^{(0)}&=E^{(0)}+P^{(0)}=(D+1) P^{(0)}, \\
S^{(0)}&=(D+1) P^{(0)} / T.
\end{align}
Thus, we can write Eq.~(\ref{energy-balance}) in another form
\begin{equation}
\frac{\partial S^{(0)}}{\partial t}+\left(1+D^{-1}\right) S^{(0)} \nabla \cdot \boldsymbol{u}=0.
\end{equation}
This equation can be compared with Eq.~(\ref{relative-entropy}). We see that the common drift velocity $\Vu$ plays the role of relative velocity $\bm{w}$ between the normal and super ﬂuid in helium \Rmnum{2}. But the origin of $\Vu$ is the momentum conserving normal phonon scattering, which does not need to be quantum-mechanical. Combining with Eqs.~(\ref{momentum-balance}), we get an equation for the entropy
\begin{equation}
\frac{\partial^{2} S^{(0)}}{\partial t^{2}}-\frac{v_{g}^{2}}{D} \nabla^{2} S^{(0)}=0,
\end{equation}
which gives rises to the drifting second sound with velocity
$v_{p}=v_{g} / \sqrt{D}$.

\begin{figure} 
  \centering 
   \includegraphics[scale=0.46]{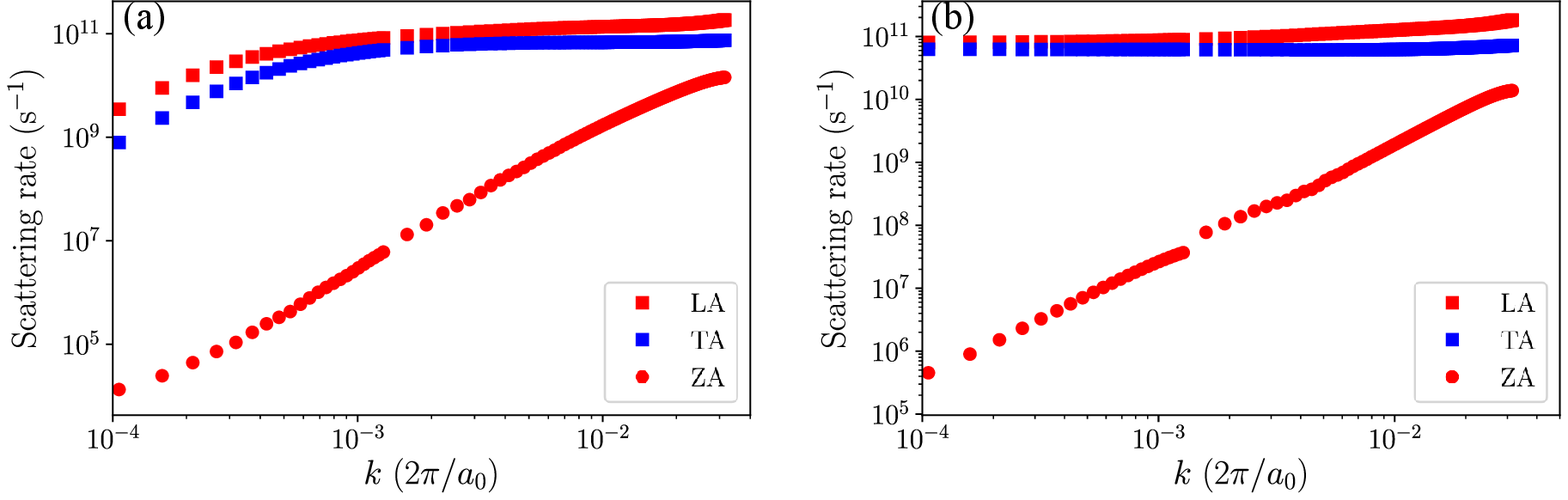}
  \caption{The scattering rate of acoustic phonon modes for graphene at 100K before (a) and after (b) the symmetrization. The ZA mode shows the smallest scattering due to mirror symmetry of graphene about the 2D plane. }
  \label{fig:scattering-rate} 
\end{figure}

\begin{figure}[h] 
  \centering 
  \includegraphics[scale=0.42]{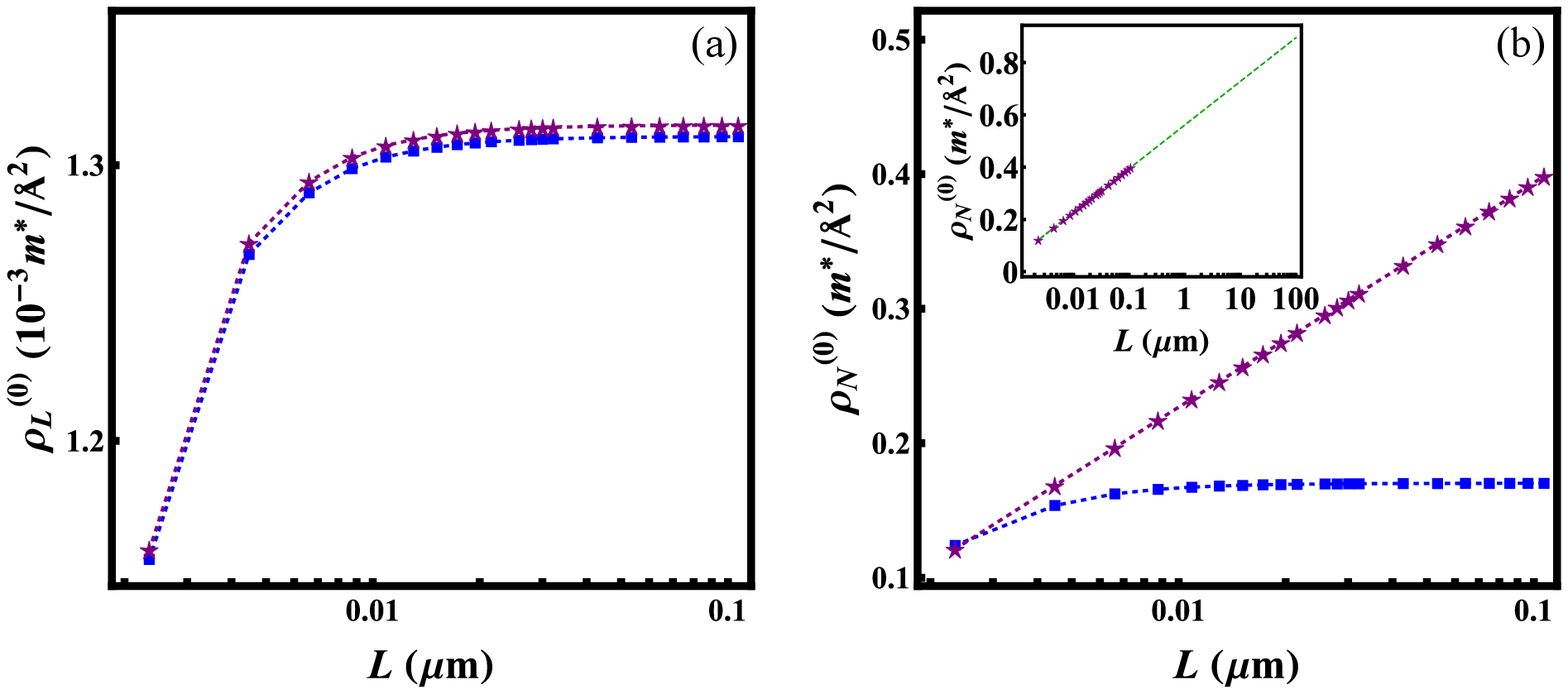} 
  \caption{Dependence of the effective mass density $\rho^{(0)}$ on the sample size for linear (a) and flexural (b) phonon modes of graphene at 100 K. Results before and after symmetrization of the forces constants are shown in blue, square and purple, star symbols, respectively. The logarithmic dependence of $\rho_N^{(0)}$ on $L$ is recovered after the symmetrization. Note that $\rho_L^{(0)}$ is two orders of magnitude smaller than $\rho_N^{(0)}$. The inset in (b) shows fitting of $\rho_N^{(0)}$ to ${\rm ln}L$. Extrapolation of $\rho_N^{(0)}$ to $L=50$ $\mu$m is used in the numerical calculation of Fig.~\ref{fig:w-k}. $m^* = \hbar/(2a)$ is the effective mass defined in the main text. }
  \label{fig:rho-all} 
\end{figure}

\section{Details of the DFT calculation}
\label{app:dft}
Phonon dispersion relation of graphene is calculated using Vienna Ab-initio Simulation Package (VASP) \cite{PhysRevB.54.11169,KRESSE199615} combined with phonopy
(Fig.~\ref{fig:wk-com})\cite{phonopy}. 
%(Fig.~\ref{fig:wk-com} and Fig.~\ref{fig:wk-com-BN} )\cite{phonopy}. 
We use a supercell size $5\times 5 \times 1$ to calculate the second order force constants. The numerical flexural phonon dispersion does not follow the ideal quadratic relation near $\Gamma$ (blue dashed dotted line). Additional symmetrization is used to recover the quadratic dispersion (red solid line)\cite{Jes2016Physically}. This slight change of dispersion has important influence on the second sound dispersion (Fig.~\ref{fig:w-k}).

To obtain the relaxation times, we calculate the third order force constant using phono3py\cite{phono3py} with the same supercell size. 
The numbers of mesh points for reciprocal space sampling are $501\times 501\times 1$. The scattering rates before and after the symmetrization of graphene are shown in Fig.~\ref{fig:scattering-rate}. %for graphene and Fig.~\ref{fig:scattering-rate-si} for silicene, respectively. 
The viscosity, thermal conductivity under relaxation time approximation  and the effective mass density are obtained using these numerical results and used to calculate the second sound dispersion. As an example, we show in Fig.~\ref{fig:rho-all} the dependence of $\rho_L^{(0)}$ (a) and $\rho_N^{(0)}$ (b) on the sample size before and after the symmetrization.

\bibliography{apssamp,extra,jthydro,Appendix,divergence}

\providecommand{\noopsort}[1]{}\providecommand{\singleletter}[1]{#1}%
\begin{thebibliography}{77}%
\makeatletter
\providecommand \@ifxundefined [1]{%
 \@ifx{#1\undefined}
}%
\providecommand \@ifnum [1]{%
 \ifnum #1\expandafter \@firstoftwo
 \else \expandafter \@secondoftwo
 \fi
}%
\providecommand \@ifx [1]{%
 \ifx #1\expandafter \@firstoftwo
 \else \expandafter \@secondoftwo
 \fi
}%
\providecommand \natexlab [1]{#1}%
\providecommand \enquote  [1]{``#1''}%
\providecommand \bibnamefont  [1]{#1}%
\providecommand \bibfnamefont [1]{#1}%
\providecommand \citenamefont [1]{#1}%
\providecommand \href@noop [0]{\@secondoftwo}%
\providecommand \href [0]{\begingroup \@sanitize@url \@href}%
\providecommand \@href[1]{\@@startlink{#1}\@@href}%
\providecommand \@@href[1]{\endgroup#1\@@endlink}%
\providecommand \@sanitize@url [0]{\catcode `\\12\catcode `\$12\catcode
  `\&12\catcode `\#12\catcode `\^12\catcode `\_12\catcode `\%12\relax}%
\providecommand \@@startlink[1]{}%
\providecommand \@@endlink[0]{}%
\providecommand \url  [0]{\begingroup\@sanitize@url \@url }%
\providecommand \@url [1]{\endgroup\@href {#1}{\urlprefix }}%
\providecommand \urlprefix  [0]{URL }%
\providecommand \Eprint [0]{\href }%
\providecommand \doibase [0]{https://doi.org/}%
\providecommand \selectlanguage [0]{\@gobble}%
\providecommand \bibinfo  [0]{\@secondoftwo}%
\providecommand \bibfield  [0]{\@secondoftwo}%
\providecommand \translation [1]{[#1]}%
\providecommand \BibitemOpen [0]{}%
\providecommand \bibitemStop [0]{}%
\providecommand \bibitemNoStop [0]{.\EOS\space}%
\providecommand \EOS [0]{\spacefactor3000\relax}%
\providecommand \BibitemShut  [1]{\csname bibitem#1\endcsname}%
\let\auto@bib@innerbib\@empty
%</preamble>
\bibitem [{\citenamefont {Chen}(2021)}]{Chen2021}%
  \BibitemOpen
  \bibfield  {author} {\bibinfo {author} {\bibfnamefont {G.}~\bibnamefont
  {Chen}},\ }\bibfield  {title} {\bibinfo {title} {Non-fourier phonon heat
  conduction at the microscale and nanoscale},\ }\href
  {https://doi.org/10.1038/s42254-021-00334-1} {\bibfield  {journal} {\bibinfo
  {journal} {Nat. Rev. Phys.}\ }\textbf {\bibinfo {volume} {3}},\ \bibinfo
  {pages} {555} (\bibinfo {year} {2021})}\BibitemShut {NoStop}%
\bibitem [{\citenamefont {Wang}\ \emph {et~al.}(2008)\citenamefont {Wang},
  \citenamefont {Wang},\ and\ \citenamefont {Lü}}]{Wang2008}%
  \BibitemOpen
  \bibfield  {author} {\bibinfo {author} {\bibfnamefont {J.-S.}\ \bibnamefont
  {Wang}}, \bibinfo {author} {\bibfnamefont {J.}~\bibnamefont {Wang}},\ and\
  \bibinfo {author} {\bibfnamefont {J.~T.}\ \bibnamefont {Lü}},\ }\bibfield
  {title} {\bibinfo {title} {Quantum thermal transport in nanostructures},\
  }\href {https://doi.org/10.1140/epjb/e2008-00195-8} {\bibfield  {journal}
  {\bibinfo  {journal} {Eur. Phys. J. B}\ }\textbf {\bibinfo {volume} {62}},\
  \bibinfo {pages} {381} (\bibinfo {year} {2008})}\BibitemShut {NoStop}%
\bibitem [{\citenamefont {Dhar}(2008)}]{Dhar2008}%
  \BibitemOpen
  \bibfield  {author} {\bibinfo {author} {\bibfnamefont {A.}~\bibnamefont
  {Dhar}},\ }\bibfield  {title} {\bibinfo {title} {Heat transport in
  low-dimensional systems},\ }\href {https://doi.org/10.1080/00018730802538522}
  {\bibfield  {journal} {\bibinfo  {journal} {Adv. Phys.}\ }\textbf {\bibinfo
  {volume} {57}},\ \bibinfo {pages} {457} (\bibinfo {year} {2008})}\BibitemShut
  {NoStop}%
\bibitem [{\citenamefont {Li}\ \emph {et~al.}(2012)\citenamefont {Li},
  \citenamefont {Ren}, \citenamefont {Wang}, \citenamefont {Zhang},
  \citenamefont {H\"anggi},\ and\ \citenamefont {Li}}]{Nianbei-RMP}%
  \BibitemOpen
  \bibfield  {author} {\bibinfo {author} {\bibfnamefont {N.}~\bibnamefont
  {Li}}, \bibinfo {author} {\bibfnamefont {J.}~\bibnamefont {Ren}}, \bibinfo
  {author} {\bibfnamefont {L.}~\bibnamefont {Wang}}, \bibinfo {author}
  {\bibfnamefont {G.}~\bibnamefont {Zhang}}, \bibinfo {author} {\bibfnamefont
  {P.}~\bibnamefont {H\"anggi}},\ and\ \bibinfo {author} {\bibfnamefont
  {B.}~\bibnamefont {Li}},\ }\bibfield  {title} {\bibinfo {title} {Colloquium:
  Phononics: Manipulating heat flow with electronic analogs and beyond},\
  }\href {https://doi.org/10.1103/RevModPhys.84.1045} {\bibfield  {journal}
  {\bibinfo  {journal} {Rev. Mod. Phys.}\ }\textbf {\bibinfo {volume} {84}},\
  \bibinfo {pages} {1045} (\bibinfo {year} {2012})}\BibitemShut {NoStop}%
\bibitem [{\citenamefont {Gu}\ \emph {et~al.}(2018)\citenamefont {Gu},
  \citenamefont {Wei}, \citenamefont {Yin}, \citenamefont {Li},\ and\
  \citenamefont {Yang}}]{gu_phononic_2018}%
  \BibitemOpen
  \bibfield  {author} {\bibinfo {author} {\bibfnamefont {X.}~\bibnamefont
  {Gu}}, \bibinfo {author} {\bibfnamefont {Y.}~\bibnamefont {Wei}}, \bibinfo
  {author} {\bibfnamefont {X.}~\bibnamefont {Yin}}, \bibinfo {author}
  {\bibfnamefont {B.}~\bibnamefont {Li}},\ and\ \bibinfo {author}
  {\bibfnamefont {R.}~\bibnamefont {Yang}},\ }\bibfield  {title} {\bibinfo
  {title} {Colloquium: Phononic thermal properties of two-dimensional
  materials},\ }\href {https://doi.org/10.1103/RevModPhys.90.041002} {\bibfield
   {journal} {\bibinfo  {journal} {Rev. Mod. Phys.}\ }\textbf {\bibinfo
  {volume} {90}},\ \bibinfo {pages} {041002} (\bibinfo {year}
  {2018})}\BibitemShut {NoStop}%
\bibitem [{\citenamefont {Lepri}\ \emph {et~al.}(2003)\citenamefont {Lepri},
  \citenamefont {Livi},\ and\ \citenamefont {Politi}}]{lepri_thermal_2003}%
  \BibitemOpen
  \bibfield  {author} {\bibinfo {author} {\bibfnamefont {S.}~\bibnamefont
  {Lepri}}, \bibinfo {author} {\bibfnamefont {R.}~\bibnamefont {Livi}},\ and\
  \bibinfo {author} {\bibfnamefont {A.}~\bibnamefont {Politi}},\ }\bibfield
  {title} {\bibinfo {title} {Thermal conduction in classical low-dimensional
  lattices},\ }\href {https://doi.org/10.1016/S0370-1573(02)00558-6} {\bibfield
   {journal} {\bibinfo  {journal} {Phys. Rep.}\ }\textbf {\bibinfo {volume}
  {377}},\ \bibinfo {pages} {1} (\bibinfo {year} {2003})}\BibitemShut {NoStop}%
\bibitem [{\citenamefont {Beck}\ \emph {et~al.}(1974)\citenamefont {Beck},
  \citenamefont {Meier},\ and\ \citenamefont {Thellung}}]{beck_phonon_1974}%
  \BibitemOpen
  \bibfield  {author} {\bibinfo {author} {\bibfnamefont {H.}~\bibnamefont
  {Beck}}, \bibinfo {author} {\bibfnamefont {P.~F.}\ \bibnamefont {Meier}},\
  and\ \bibinfo {author} {\bibfnamefont {A.}~\bibnamefont {Thellung}},\
  }\bibfield  {title} {\bibinfo {title} {Phonon hydrodynamics in solids},\
  }\href {https://doi.org/10.1002/pssa.2210240102} {\bibfield  {journal}
  {\bibinfo  {journal} {Phys. Stat. Sol. (a)}\ }\textbf {\bibinfo {volume}
  {24}},\ \bibinfo {pages} {11} (\bibinfo {year} {1974})}\BibitemShut {NoStop}%
\bibitem [{\citenamefont {Joseph}\ and\ \citenamefont
  {Preziosi}(1989)}]{joseph_heat_1989-1}%
  \BibitemOpen
  \bibfield  {author} {\bibinfo {author} {\bibfnamefont {D.~D.}\ \bibnamefont
  {Joseph}}\ and\ \bibinfo {author} {\bibfnamefont {L.}~\bibnamefont
  {Preziosi}},\ }\bibfield  {title} {\bibinfo {title} {Heat waves},\ }\href
  {https://doi.org/10.1103/RevModPhys.61.41} {\bibfield  {journal} {\bibinfo
  {journal} {Rev. Mod. Phys.}\ }\textbf {\bibinfo {volume} {61}},\ \bibinfo
  {pages} {41} (\bibinfo {year} {1989})}\BibitemShut {NoStop}%
\bibitem [{\citenamefont {Lee}\ and\ \citenamefont {Li}(2020)}]{Lee-Review}%
  \BibitemOpen
  \bibfield  {author} {\bibinfo {author} {\bibfnamefont {S.}~\bibnamefont
  {Lee}}\ and\ \bibinfo {author} {\bibfnamefont {X.}~\bibnamefont {Li}},\
  }\bibfield  {title} {\bibinfo {title} {Hydrodynamic phonon transport: past,
  present and prospects},\ }in\ \href
  {https://doi.org/10.1088/978-0-7503-1738-2ch1} {\emph {\bibinfo {booktitle}
  {Nanoscale Energy Transport}}},\ \bibinfo {series and number} {2053-2563}\
  (\bibinfo  {publisher} {IOP Publishing},\ \bibinfo {year} {2020})\ pp.\
  \bibinfo {pages} {1--1 to 1--26}\BibitemShut {NoStop}%
\bibitem [{\citenamefont {Wang}\ \emph {et~al.}(2020)\citenamefont {Wang},
  \citenamefont {Dai},\ and\ \citenamefont {Huang}}]{wang2020}%
  \BibitemOpen
  \bibfield  {author} {\bibinfo {author} {\bibfnamefont {J.}~\bibnamefont
  {Wang}}, \bibinfo {author} {\bibfnamefont {G.}~\bibnamefont {Dai}},\ and\
  \bibinfo {author} {\bibfnamefont {J.}~\bibnamefont {Huang}},\ }\bibfield
  {title} {\bibinfo {title} {Thermal metamaterial: Fundamental, application,
  and outlook},\ }\href {https://doi.org/10.1016/j.isci.2020.101637} {\bibfield
   {journal} {\bibinfo  {journal} {iScience}\ }\textbf {\bibinfo {volume}
  {23}},\ \bibinfo {pages} {101637} (\bibinfo {year} {2020})}\BibitemShut
  {NoStop}%
\bibitem [{\citenamefont {Nakamura}(2019)}]{Nakamura2019}%
  \BibitemOpen
  \bibfield  {author} {\bibinfo {author} {\bibfnamefont {K.}~\bibnamefont
  {Nakamura}},\ }\href@noop {} {\emph {\bibinfo {title} {Quantum phononics:
  introduction to ultrafast dynamics of optical phonons}}}\ (\bibinfo
  {publisher} {Cham, Switzerland:Springer},\ \bibinfo {year}
  {2019})\BibitemShut {NoStop}%
\bibitem [{\citenamefont {Li}\ \emph {et~al.}(2021)\citenamefont {Li},
  \citenamefont {Li}, \citenamefont {Han}, \citenamefont {Zheng}, \citenamefont
  {Li}, \citenamefont {Li}, \citenamefont {Fan},\ and\ \citenamefont
  {Qiu}}]{li_transforming_2021}%
  \BibitemOpen
  \bibfield  {author} {\bibinfo {author} {\bibfnamefont {Y.}~\bibnamefont
  {Li}}, \bibinfo {author} {\bibfnamefont {W.}~\bibnamefont {Li}}, \bibinfo
  {author} {\bibfnamefont {T.}~\bibnamefont {Han}}, \bibinfo {author}
  {\bibfnamefont {X.}~\bibnamefont {Zheng}}, \bibinfo {author} {\bibfnamefont
  {J.}~\bibnamefont {Li}}, \bibinfo {author} {\bibfnamefont {B.}~\bibnamefont
  {Li}}, \bibinfo {author} {\bibfnamefont {S.}~\bibnamefont {Fan}},\ and\
  \bibinfo {author} {\bibfnamefont {C.-W.}\ \bibnamefont {Qiu}},\ }\bibfield
  {title} {\bibinfo {title} {Transforming heat transfer with thermal
  metamaterials and devices},\ }\href
  {https://doi.org/10.1038/s41578-021-00283-2} {\bibfield  {journal} {\bibinfo
  {journal} {Nat. Rev. Mater.}\ }\textbf {\bibinfo {volume} {6}},\ \bibinfo
  {pages} {488} (\bibinfo {year} {2021})}\BibitemShut {NoStop}%
\bibitem [{\citenamefont {Ward}\ and\ \citenamefont
  {Wilks}(1951)}]{ward_velocity_1951}%
  \BibitemOpen
  \bibfield  {author} {\bibinfo {author} {\bibfnamefont {J.~C.}\ \bibnamefont
  {Ward}}\ and\ \bibinfo {author} {\bibfnamefont {J.}~\bibnamefont {Wilks}},\
  }\bibfield  {title} {\bibinfo {title} {The velocity of second sound in liquid
  helium near the absolute zero},\ }\href
  {https://doi.org/10.1080/14786445108561271} {\bibfield  {journal} {\bibinfo
  {journal} {Phil. Mag.}\ }\textbf {\bibinfo {volume} {42}},\ \bibinfo {pages}
  {314} (\bibinfo {year} {1951})}\BibitemShut {NoStop}%
\bibitem [{\citenamefont {Ward}\ and\ \citenamefont
  {Wilks}(1952)}]{ward_iii_1952}%
  \BibitemOpen
  \bibfield  {author} {\bibinfo {author} {\bibfnamefont {J.~C.}\ \bibnamefont
  {Ward}}\ and\ \bibinfo {author} {\bibfnamefont {J.}~\bibnamefont {Wilks}},\
  }\bibfield  {title} {\bibinfo {title} {{Second} sound and the
  thermo-mechanical effect at very low temperatures},\ }\href
  {https://doi.org/10.1080/14786440108520965} {\bibfield  {journal} {\bibinfo
  {journal} {Phil. Mag.}\ }\textbf {\bibinfo {volume} {43}},\ \bibinfo {pages}
  {48} (\bibinfo {year} {1952})}\BibitemShut {NoStop}%
\bibitem [{\citenamefont {Sussmann}\ and\ \citenamefont
  {Thellung}(1963)}]{sussmann_thermal_1963}%
  \BibitemOpen
  \bibfield  {author} {\bibinfo {author} {\bibfnamefont {J.~A.}\ \bibnamefont
  {Sussmann}}\ and\ \bibinfo {author} {\bibfnamefont {A.}~\bibnamefont
  {Thellung}},\ }\bibfield  {title} {\bibinfo {title} {Thermal conductivity of
  perfect dielectric crystals in the absence of umklapp processes},\ }\href
  {https://doi.org/10.1088/0370-1328/81/6/318} {\bibfield  {journal} {\bibinfo
  {journal} {Proc. Phys. Soc.}\ }\textbf {\bibinfo {volume} {81}},\ \bibinfo
  {pages} {1122} (\bibinfo {year} {1963})}\BibitemShut {NoStop}%
\bibitem [{\citenamefont {Gurzhi}(1968)}]{gurzhi_hydrodynamic_1968}%
  \BibitemOpen
  \bibfield  {author} {\bibinfo {author} {\bibfnamefont {R.~N.}\ \bibnamefont
  {Gurzhi}},\ }\bibfield  {title} {\bibinfo {title} {Hydrodynamic effects in
  solids at low temperature},\ }\href
  {https://doi.org/10.1070/PU1968v011n02ABEH003815} {\bibfield  {journal}
  {\bibinfo  {journal} {Sov. Phys. Usp.}\ }\textbf {\bibinfo {volume} {11}},\
  \bibinfo {pages} {255} (\bibinfo {year} {1968})}\BibitemShut {NoStop}%
\bibitem [{\citenamefont {Guyer}\ and\ \citenamefont
  {Krumhansl}(1966{\natexlab{a}})}]{guyer_solution_1966}%
  \BibitemOpen
  \bibfield  {author} {\bibinfo {author} {\bibfnamefont {R.~A.}\ \bibnamefont
  {Guyer}}\ and\ \bibinfo {author} {\bibfnamefont {J.~A.}\ \bibnamefont
  {Krumhansl}},\ }\bibfield  {title} {\bibinfo {title} {Solution of the
  {Linearized} {Phonon} {Boltzmann} {Equation}},\ }\href
  {https://doi.org/10.1103/PhysRev.148.766} {\bibfield  {journal} {\bibinfo
  {journal} {Phys. Rev.}\ }\textbf {\bibinfo {volume} {148}},\ \bibinfo {pages}
  {766} (\bibinfo {year} {1966}{\natexlab{a}})}\BibitemShut {NoStop}%
\bibitem [{\citenamefont {Guyer}\ and\ \citenamefont
  {Krumhansl}(1966{\natexlab{b}})}]{guyer_thermal_1966}%
  \BibitemOpen
  \bibfield  {author} {\bibinfo {author} {\bibfnamefont {R.~A.}\ \bibnamefont
  {Guyer}}\ and\ \bibinfo {author} {\bibfnamefont {J.~A.}\ \bibnamefont
  {Krumhansl}},\ }\bibfield  {title} {\bibinfo {title} {Thermal {Conductivity},
  {Second} {Sound}, and {Phonon} {Hydrodynamic} {Phenomena} in {Nonmetallic}
  {Crystals}},\ }\href {https://doi.org/10.1103/PhysRev.148.778} {\bibfield
  {journal} {\bibinfo  {journal} {Phys. Rev.}\ }\textbf {\bibinfo {volume}
  {148}},\ \bibinfo {pages} {778} (\bibinfo {year}
  {1966}{\natexlab{b}})}\BibitemShut {NoStop}%
\bibitem [{\citenamefont {Hardy}(1970)}]{hardy_phonon_1970}%
  \BibitemOpen
  \bibfield  {author} {\bibinfo {author} {\bibfnamefont {R.~J.}\ \bibnamefont
  {Hardy}},\ }\bibfield  {title} {\bibinfo {title} {Phonon {Boltzmann}
  {Equation} and {Second} {Sound} in {Solids}},\ }\href
  {https://doi.org/10.1103/PhysRevB.2.1193} {\bibfield  {journal} {\bibinfo
  {journal} {Phys. Rev. B}\ }\textbf {\bibinfo {volume} {2}},\ \bibinfo {pages}
  {1193} (\bibinfo {year} {1970})}\BibitemShut {NoStop}%
\bibitem [{\citenamefont {Ackerman}\ \emph {et~al.}(1966)\citenamefont
  {Ackerman}, \citenamefont {Bertman}, \citenamefont {Fairbank},\ and\
  \citenamefont {Guyer}}]{ackerman_second_1966}%
  \BibitemOpen
  \bibfield  {author} {\bibinfo {author} {\bibfnamefont {C.~C.}\ \bibnamefont
  {Ackerman}}, \bibinfo {author} {\bibfnamefont {B.}~\bibnamefont {Bertman}},
  \bibinfo {author} {\bibfnamefont {H.~A.}\ \bibnamefont {Fairbank}},\ and\
  \bibinfo {author} {\bibfnamefont {R.~A.}\ \bibnamefont {Guyer}},\ }\bibfield
  {title} {\bibinfo {title} {Second sound in solid helium},\ }\href
  {https://doi.org/10.1103/PhysRevLett.16.789} {\bibfield  {journal} {\bibinfo
  {journal} {Phys. Rev. Lett.}\ }\textbf {\bibinfo {volume} {16}},\ \bibinfo
  {pages} {789} (\bibinfo {year} {1966})}\BibitemShut {NoStop}%
\bibitem [{\citenamefont {McNelly}\ \emph {et~al.}(1970)\citenamefont
  {McNelly}, \citenamefont {Rogers}, \citenamefont {Channin}, \citenamefont
  {Rollefson}, \citenamefont {Goubau}, \citenamefont {Schmidt}, \citenamefont
  {Krumhansl},\ and\ \citenamefont {Pohl}}]{mcnelly_heat_1970}%
  \BibitemOpen
  \bibfield  {author} {\bibinfo {author} {\bibfnamefont {T.~F.}\ \bibnamefont
  {McNelly}}, \bibinfo {author} {\bibfnamefont {S.~J.}\ \bibnamefont {Rogers}},
  \bibinfo {author} {\bibfnamefont {D.~J.}\ \bibnamefont {Channin}}, \bibinfo
  {author} {\bibfnamefont {R.~J.}\ \bibnamefont {Rollefson}}, \bibinfo {author}
  {\bibfnamefont {W.~M.}\ \bibnamefont {Goubau}}, \bibinfo {author}
  {\bibfnamefont {G.~E.}\ \bibnamefont {Schmidt}}, \bibinfo {author}
  {\bibfnamefont {J.~A.}\ \bibnamefont {Krumhansl}},\ and\ \bibinfo {author}
  {\bibfnamefont {R.~O.}\ \bibnamefont {Pohl}},\ }\bibfield  {title} {\bibinfo
  {title} {Heat pulses in naf: Onset of second sound},\ }\href
  {https://doi.org/10.1103/PhysRevLett.24.100} {\bibfield  {journal} {\bibinfo
  {journal} {Phys. Rev. Lett.}\ }\textbf {\bibinfo {volume} {24}},\ \bibinfo
  {pages} {100} (\bibinfo {year} {1970})}\BibitemShut {NoStop}%
\bibitem [{\citenamefont {Narayanamurti}\ and\ \citenamefont
  {Dynes}(1972)}]{narayanamurti_observation_1972}%
  \BibitemOpen
  \bibfield  {author} {\bibinfo {author} {\bibfnamefont {V.}~\bibnamefont
  {Narayanamurti}}\ and\ \bibinfo {author} {\bibfnamefont {R.~C.}\ \bibnamefont
  {Dynes}},\ }\bibfield  {title} {\bibinfo {title} {Observation of second sound
  in bismuth},\ }\href {https://doi.org/10.1103/PhysRevLett.28.1461} {\bibfield
   {journal} {\bibinfo  {journal} {Phys. Rev. Lett.}\ }\textbf {\bibinfo
  {volume} {28}},\ \bibinfo {pages} {1461} (\bibinfo {year}
  {1972})}\BibitemShut {NoStop}%
\bibitem [{\citenamefont {Koreeda}\ \emph {et~al.}(2007)\citenamefont
  {Koreeda}, \citenamefont {Takano},\ and\ \citenamefont
  {Saikan}}]{koreeda_second_2007}%
  \BibitemOpen
  \bibfield  {author} {\bibinfo {author} {\bibfnamefont {A.}~\bibnamefont
  {Koreeda}}, \bibinfo {author} {\bibfnamefont {R.}~\bibnamefont {Takano}},\
  and\ \bibinfo {author} {\bibfnamefont {S.}~\bibnamefont {Saikan}},\
  }\bibfield  {title} {\bibinfo {title} {Second sound in
  ${\mathrm{srtio}}_{3}$},\ }\href
  {https://doi.org/10.1103/PhysRevLett.99.265502} {\bibfield  {journal}
  {\bibinfo  {journal} {Phys. Rev. Lett.}\ }\textbf {\bibinfo {volume} {99}},\
  \bibinfo {pages} {265502} (\bibinfo {year} {2007})}\BibitemShut {NoStop}%
\bibitem [{\citenamefont {Lee}\ \emph {et~al.}(2015)\citenamefont {Lee},
  \citenamefont {Broido}, \citenamefont {Esfarjani},\ and\ \citenamefont
  {Chen}}]{lee_hydrodynamic_2015}%
  \BibitemOpen
  \bibfield  {author} {\bibinfo {author} {\bibfnamefont {S.}~\bibnamefont
  {Lee}}, \bibinfo {author} {\bibfnamefont {D.}~\bibnamefont {Broido}},
  \bibinfo {author} {\bibfnamefont {K.}~\bibnamefont {Esfarjani}},\ and\
  \bibinfo {author} {\bibfnamefont {G.}~\bibnamefont {Chen}},\ }\bibfield
  {title} {\bibinfo {title} {Hydrodynamic phonon transport in suspended
  graphene},\ }\href {https://doi.org/10.1038/ncomms7290} {\bibfield  {journal}
  {\bibinfo  {journal} {Nat. Commun.}\ }\textbf {\bibinfo {volume} {6}},\
  \bibinfo {pages} {6290} (\bibinfo {year} {2015})}\BibitemShut {NoStop}%
\bibitem [{\citenamefont {Cepellotti}\ \emph {et~al.}(2015)\citenamefont
  {Cepellotti}, \citenamefont {Fugallo}, \citenamefont {Paulatto},
  \citenamefont {Lazzeri}, \citenamefont {Mauri},\ and\ \citenamefont
  {Marzari}}]{cepellotti_phonon_2015}%
  \BibitemOpen
  \bibfield  {author} {\bibinfo {author} {\bibfnamefont {A.}~\bibnamefont
  {Cepellotti}}, \bibinfo {author} {\bibfnamefont {G.}~\bibnamefont {Fugallo}},
  \bibinfo {author} {\bibfnamefont {L.}~\bibnamefont {Paulatto}}, \bibinfo
  {author} {\bibfnamefont {M.}~\bibnamefont {Lazzeri}}, \bibinfo {author}
  {\bibfnamefont {F.}~\bibnamefont {Mauri}},\ and\ \bibinfo {author}
  {\bibfnamefont {N.}~\bibnamefont {Marzari}},\ }\bibfield  {title} {\bibinfo
  {title} {Phonon hydrodynamics in two-dimensional materials},\ }\href
  {https://doi.org/10.1038/ncomms7400} {\bibfield  {journal} {\bibinfo
  {journal} {Nat. Commun.}\ }\textbf {\bibinfo {volume} {6}},\ \bibinfo {pages}
  {6400} (\bibinfo {year} {2015})}\BibitemShut {NoStop}%
\bibitem [{\citenamefont {Ding}\ \emph {et~al.}(2018)\citenamefont {Ding},
  \citenamefont {Zhou}, \citenamefont {Song}, \citenamefont {Chiloyan},
  \citenamefont {Li}, \citenamefont {Liu},\ and\ \citenamefont
  {Chen}}]{ding_phonon_2017}%
  \BibitemOpen
  \bibfield  {author} {\bibinfo {author} {\bibfnamefont {Z.}~\bibnamefont
  {Ding}}, \bibinfo {author} {\bibfnamefont {J.}~\bibnamefont {Zhou}}, \bibinfo
  {author} {\bibfnamefont {B.}~\bibnamefont {Song}}, \bibinfo {author}
  {\bibfnamefont {V.}~\bibnamefont {Chiloyan}}, \bibinfo {author}
  {\bibfnamefont {M.}~\bibnamefont {Li}}, \bibinfo {author} {\bibfnamefont
  {T.-H.}\ \bibnamefont {Liu}},\ and\ \bibinfo {author} {\bibfnamefont
  {G.}~\bibnamefont {Chen}},\ }\bibfield  {title} {\bibinfo {title} {Phonon
  {Hydrodynamic} {Heat} {Conduction} and {Knudsen} {Minimum} in {Graphite}},\
  }\href {https://doi.org/10.1021/acs.nanolett.7b04932} {\bibfield  {journal}
  {\bibinfo  {journal} {Nano Lett.}\ }\textbf {\bibinfo {volume} {18}},\
  \bibinfo {pages} {638} (\bibinfo {year} {2018})}\BibitemShut {NoStop}%
\bibitem [{\citenamefont {Martelli}\ \emph {et~al.}(2018)\citenamefont
  {Martelli}, \citenamefont {Jiménez}, \citenamefont {Continentino},
  \citenamefont {Baggio-Saitovitch},\ and\ \citenamefont
  {Behnia}}]{martelli_thermal_2018}%
  \BibitemOpen
  \bibfield  {author} {\bibinfo {author} {\bibfnamefont {V.}~\bibnamefont
  {Martelli}}, \bibinfo {author} {\bibfnamefont {J.~L.}\ \bibnamefont
  {Jiménez}}, \bibinfo {author} {\bibfnamefont {M.}~\bibnamefont
  {Continentino}}, \bibinfo {author} {\bibfnamefont {E.}~\bibnamefont
  {Baggio-Saitovitch}},\ and\ \bibinfo {author} {\bibfnamefont
  {K.}~\bibnamefont {Behnia}},\ }\bibfield  {title} {\bibinfo {title} {Thermal
  {Transport} and {Phonon} {Hydrodynamics} in {Strontium} {Titanate}},\ }\href
  {https://doi.org/10.1103/PhysRevLett.120.125901} {\bibfield  {journal}
  {\bibinfo  {journal} {Phys. Rev. Lett.}\ }\textbf {\bibinfo {volume} {120}},\
  \bibinfo {pages} {125901} (\bibinfo {year} {2018})}\BibitemShut {NoStop}%
\bibitem [{\citenamefont {Machida}\ \emph {et~al.}(2018)\citenamefont
  {Machida}, \citenamefont {Subedi}, \citenamefont {Akiba}, \citenamefont
  {Miyake}, \citenamefont {Tokunaga}, \citenamefont {Akahama}, \citenamefont
  {Izawa},\ and\ \citenamefont {Behnia}}]{machida_observation_2018}%
  \BibitemOpen
  \bibfield  {author} {\bibinfo {author} {\bibfnamefont {Y.}~\bibnamefont
  {Machida}}, \bibinfo {author} {\bibfnamefont {A.}~\bibnamefont {Subedi}},
  \bibinfo {author} {\bibfnamefont {K.}~\bibnamefont {Akiba}}, \bibinfo
  {author} {\bibfnamefont {A.}~\bibnamefont {Miyake}}, \bibinfo {author}
  {\bibfnamefont {M.}~\bibnamefont {Tokunaga}}, \bibinfo {author}
  {\bibfnamefont {Y.}~\bibnamefont {Akahama}}, \bibinfo {author} {\bibfnamefont
  {K.}~\bibnamefont {Izawa}},\ and\ \bibinfo {author} {\bibfnamefont
  {K.}~\bibnamefont {Behnia}},\ }\bibfield  {title} {\bibinfo {title}
  {Observation of {Poiseuille} flow of phonons in black phosphorus},\ }\href
  {https://doi.org/10.1126/sciadv.aat3374} {\bibfield  {journal} {\bibinfo
  {journal} {Sci. Adv}\ }\textbf {\bibinfo {volume} {4}},\ \bibinfo {pages}
  {eaat3374} (\bibinfo {year} {2018})}\BibitemShut {NoStop}%
\bibitem [{\citenamefont {Machida}\ \emph {et~al.}(2020)\citenamefont
  {Machida}, \citenamefont {Matsumoto}, \citenamefont {Isono},\ and\
  \citenamefont {Behnia}}]{machida_phonon_2020}%
  \BibitemOpen
  \bibfield  {author} {\bibinfo {author} {\bibfnamefont {Y.}~\bibnamefont
  {Machida}}, \bibinfo {author} {\bibfnamefont {N.}~\bibnamefont {Matsumoto}},
  \bibinfo {author} {\bibfnamefont {T.}~\bibnamefont {Isono}},\ and\ \bibinfo
  {author} {\bibfnamefont {K.}~\bibnamefont {Behnia}},\ }\bibfield  {title}
  {\bibinfo {title} {Phonon hydrodynamics and ultrahigh–room-temperature
  thermal conductivity in thin graphite},\ }\href
  {https://doi.org/10.1126/science.aaz8043} {\bibfield  {journal} {\bibinfo
  {journal} {Science}\ }\textbf {\bibinfo {volume} {367}},\ \bibinfo {pages}
  {309} (\bibinfo {year} {2020})}\BibitemShut {NoStop}%
\bibitem [{\citenamefont {Huberman}\ \emph {et~al.}(2019)\citenamefont
  {Huberman}, \citenamefont {Duncan}, \citenamefont {Chen}, \citenamefont
  {Song}, \citenamefont {Chiloyan}, \citenamefont {Ding}, \citenamefont
  {Maznev}, \citenamefont {Chen},\ and\ \citenamefont
  {Nelson}}]{huberman_observation_2019}%
  \BibitemOpen
  \bibfield  {author} {\bibinfo {author} {\bibfnamefont {S.}~\bibnamefont
  {Huberman}}, \bibinfo {author} {\bibfnamefont {R.~A.}\ \bibnamefont
  {Duncan}}, \bibinfo {author} {\bibfnamefont {K.}~\bibnamefont {Chen}},
  \bibinfo {author} {\bibfnamefont {B.}~\bibnamefont {Song}}, \bibinfo {author}
  {\bibfnamefont {V.}~\bibnamefont {Chiloyan}}, \bibinfo {author}
  {\bibfnamefont {Z.}~\bibnamefont {Ding}}, \bibinfo {author} {\bibfnamefont
  {A.~A.}\ \bibnamefont {Maznev}}, \bibinfo {author} {\bibfnamefont
  {G.}~\bibnamefont {Chen}},\ and\ \bibinfo {author} {\bibfnamefont {K.~A.}\
  \bibnamefont {Nelson}},\ }\bibfield  {title} {\bibinfo {title} {Observation
  of second sound in graphite at temperatures above 100 {K}},\ }\href
  {https://doi.org/10.1126/science.aav3548} {\bibfield  {journal} {\bibinfo
  {journal} {Science}\ }\textbf {\bibinfo {volume} {364}},\ \bibinfo {pages}
  {375} (\bibinfo {year} {2019})}\BibitemShut {NoStop}%
\bibitem [{\citenamefont {Cepellotti}\ and\ \citenamefont
  {Marzari}(2016)}]{cepellotti_thermal_2016}%
  \BibitemOpen
  \bibfield  {author} {\bibinfo {author} {\bibfnamefont {A.}~\bibnamefont
  {Cepellotti}}\ and\ \bibinfo {author} {\bibfnamefont {N.}~\bibnamefont
  {Marzari}},\ }\bibfield  {title} {\bibinfo {title} {Thermal {Transport} in
  {Crystals} as a {Kinetic} {Theory} of {Relaxons}},\ }\href
  {https://doi.org/10.1103/PhysRevX.6.041013} {\bibfield  {journal} {\bibinfo
  {journal} {Phys. Rev. X}\ }\textbf {\bibinfo {volume} {6}},\ \bibinfo {pages}
  {041013} (\bibinfo {year} {2016})}\BibitemShut {NoStop}%
\bibitem [{\citenamefont {Shang}\ \emph {et~al.}(2020)\citenamefont {Shang},
  \citenamefont {Zhang}, \citenamefont {Guo},\ and\ \citenamefont
  {L{\"u}}}]{shang_heat_2020}%
  \BibitemOpen
  \bibfield  {author} {\bibinfo {author} {\bibfnamefont {M.-Y.}\ \bibnamefont
  {Shang}}, \bibinfo {author} {\bibfnamefont {C.}~\bibnamefont {Zhang}},
  \bibinfo {author} {\bibfnamefont {Z.}~\bibnamefont {Guo}},\ and\ \bibinfo
  {author} {\bibfnamefont {J.-T.}\ \bibnamefont {L{\"u}}},\ }\bibfield  {title}
  {\bibinfo {title} {Heat vortex in hydrodynamic phonon transport of
  two-dimensional materials},\ }\href
  {https://doi.org/10.1038/s41598-020-65221-8} {\bibfield  {journal} {\bibinfo
  {journal} {Sci. Rep.}\ }\textbf {\bibinfo {volume} {10}},\ \bibinfo {pages}
  {8272} (\bibinfo {year} {2020})}\BibitemShut {NoStop}%
\bibitem [{\citenamefont {Guo}\ and\ \citenamefont
  {Wang}(2017)}]{guo_heat_2017}%
  \BibitemOpen
  \bibfield  {author} {\bibinfo {author} {\bibfnamefont {Y.}~\bibnamefont
  {Guo}}\ and\ \bibinfo {author} {\bibfnamefont {M.}~\bibnamefont {Wang}},\
  }\bibfield  {title} {\bibinfo {title} {Heat transport in two-dimensional
  materials by directly solving the phonon {Boltzmann} equation under
  {Callaway}'s dual relaxation model},\ }\href
  {https://doi.org/10.1103/PhysRevB.96.134312} {\bibfield  {journal} {\bibinfo
  {journal} {Phys. Rev. B}\ }\textbf {\bibinfo {volume} {96}},\ \bibinfo
  {pages} {134312} (\bibinfo {year} {2017})}\BibitemShut {NoStop}%
\bibitem [{\citenamefont {Lee}\ and\ \citenamefont
  {Lindsay}(2017)}]{lee_hydrodynamic_2017}%
  \BibitemOpen
  \bibfield  {author} {\bibinfo {author} {\bibfnamefont {S.}~\bibnamefont
  {Lee}}\ and\ \bibinfo {author} {\bibfnamefont {L.}~\bibnamefont {Lindsay}},\
  }\bibfield  {title} {\bibinfo {title} {Hydrodynamic phonon drift and second
  sound in a (20,20) single-wall carbon nanotube},\ }\href
  {https://doi.org/10.1103/PhysRevB.95.184304} {\bibfield  {journal} {\bibinfo
  {journal} {Phys. Rev. B}\ }\textbf {\bibinfo {volume} {95}},\ \bibinfo
  {pages} {184304} (\bibinfo {year} {2017})}\BibitemShut {NoStop}%
\bibitem [{\citenamefont {Luo}\ \emph {et~al.}(2019)\citenamefont {Luo},
  \citenamefont {Guo}, \citenamefont {Wang},\ and\ \citenamefont
  {Yi}}]{luo_direct_2019}%
  \BibitemOpen
  \bibfield  {author} {\bibinfo {author} {\bibfnamefont {X.-P.}\ \bibnamefont
  {Luo}}, \bibinfo {author} {\bibfnamefont {Y.-Y.}\ \bibnamefont {Guo}},
  \bibinfo {author} {\bibfnamefont {M.-R.}\ \bibnamefont {Wang}},\ and\
  \bibinfo {author} {\bibfnamefont {H.-L.}\ \bibnamefont {Yi}},\ }\bibfield
  {title} {\bibinfo {title} {Direct simulation of second sound in graphene by
  solving the phonon boltzmann equation via a multiscale scheme},\ }\href
  {https://doi.org/10.1103/PhysRevB.100.155401} {\bibfield  {journal} {\bibinfo
   {journal} {Phys. Rev. B}\ }\textbf {\bibinfo {volume} {100}},\ \bibinfo
  {pages} {155401} (\bibinfo {year} {2019})}\BibitemShut {NoStop}%
\bibitem [{\citenamefont {Beardo}\ \emph {et~al.}(2020)\citenamefont {Beardo},
  \citenamefont {Hennessy}, \citenamefont {Sendra}, \citenamefont {Camacho},
  \citenamefont {Myers}, \citenamefont {Bafaluy},\ and\ \citenamefont
  {Alvarez}}]{beardo_phonon_2020}%
  \BibitemOpen
  \bibfield  {author} {\bibinfo {author} {\bibfnamefont {A.}~\bibnamefont
  {Beardo}}, \bibinfo {author} {\bibfnamefont {M.~G.}\ \bibnamefont
  {Hennessy}}, \bibinfo {author} {\bibfnamefont {L.}~\bibnamefont {Sendra}},
  \bibinfo {author} {\bibfnamefont {J.}~\bibnamefont {Camacho}}, \bibinfo
  {author} {\bibfnamefont {T.~G.}\ \bibnamefont {Myers}}, \bibinfo {author}
  {\bibfnamefont {J.}~\bibnamefont {Bafaluy}},\ and\ \bibinfo {author}
  {\bibfnamefont {F.~X.}\ \bibnamefont {Alvarez}},\ }\bibfield  {title}
  {\bibinfo {title} {Phonon hydrodynamics in frequency-domain thermoreflectance
  experiments},\ }\href {https://doi.org/10.1103/PhysRevB.101.075303}
  {\bibfield  {journal} {\bibinfo  {journal} {Phys. Rev. B}\ }\textbf {\bibinfo
  {volume} {101}},\ \bibinfo {pages} {075303} (\bibinfo {year}
  {2020})}\BibitemShut {NoStop}%
\bibitem [{\citenamefont {Torres}\ \emph {et~al.}(2018)\citenamefont {Torres},
  \citenamefont {Ziabari}, \citenamefont {Torelló}, \citenamefont {Bafaluy},
  \citenamefont {Camacho}, \citenamefont {Cartoixà}, \citenamefont
  {Shakouri},\ and\ \citenamefont {Alvarez}}]{torres_emergence_2018}%
  \BibitemOpen
  \bibfield  {author} {\bibinfo {author} {\bibfnamefont {P.}~\bibnamefont
  {Torres}}, \bibinfo {author} {\bibfnamefont {A.}~\bibnamefont {Ziabari}},
  \bibinfo {author} {\bibfnamefont {A.}~\bibnamefont {Torelló}}, \bibinfo
  {author} {\bibfnamefont {J.}~\bibnamefont {Bafaluy}}, \bibinfo {author}
  {\bibfnamefont {J.}~\bibnamefont {Camacho}}, \bibinfo {author} {\bibfnamefont
  {X.}~\bibnamefont {Cartoixà}}, \bibinfo {author} {\bibfnamefont
  {A.}~\bibnamefont {Shakouri}},\ and\ \bibinfo {author} {\bibfnamefont
  {F.~X.}\ \bibnamefont {Alvarez}},\ }\bibfield  {title} {\bibinfo {title}
  {Emergence of hydrodynamic heat transport in semiconductors at the
  nanoscale},\ }\href {https://doi.org/10.1103/PhysRevMaterials.2.076001}
  {\bibfield  {journal} {\bibinfo  {journal} {Phys. Rev. Mater.}\ }\textbf
  {\bibinfo {volume} {2}},\ \bibinfo {pages} {076001} (\bibinfo {year}
  {2018})}\BibitemShut {NoStop}%
\bibitem [{\citenamefont {Zhang}\ \emph {et~al.}(2022)\citenamefont {Zhang},
  \citenamefont {Ma}, \citenamefont {Shang}, \citenamefont {Wan}, \citenamefont
  {L{\"u}}, \citenamefont {Guo}, \citenamefont {Li},\ and\ \citenamefont
  {Yang}}]{zhang2020violation}%
  \BibitemOpen
  \bibfield  {author} {\bibinfo {author} {\bibfnamefont {C.}~\bibnamefont
  {Zhang}}, \bibinfo {author} {\bibfnamefont {D.}~\bibnamefont {Ma}}, \bibinfo
  {author} {\bibfnamefont {M.}~\bibnamefont {Shang}}, \bibinfo {author}
  {\bibfnamefont {X.}~\bibnamefont {Wan}}, \bibinfo {author} {\bibfnamefont
  {J.-T.}\ \bibnamefont {L{\"u}}}, \bibinfo {author} {\bibfnamefont
  {Z.}~\bibnamefont {Guo}}, \bibinfo {author} {\bibfnamefont {B.}~\bibnamefont
  {Li}},\ and\ \bibinfo {author} {\bibfnamefont {N.}~\bibnamefont {Yang}},\
  }\bibfield  {title} {\bibinfo {title} {Graded thermal conductivity in {2D}
  and {3D} homogeneous hotspot systems},\ }\href
  {https://doi.org/https://doi.org/10.1016/j.mtphys.2022.100605} {\bibfield
  {journal} {\bibinfo  {journal} {Mater. Today Phys.}\ ,\ \bibinfo {pages}
  {100605}} (\bibinfo {year} {2022})}\BibitemShut {NoStop}%
\bibitem [{\citenamefont {Yu}\ \emph {et~al.}(2021)\citenamefont {Yu},
  \citenamefont {Ouyang},\ and\ \citenamefont {Chen}}]{yu_perspective_2021}%
  \BibitemOpen
  \bibfield  {author} {\bibinfo {author} {\bibfnamefont {C.}~\bibnamefont
  {Yu}}, \bibinfo {author} {\bibfnamefont {Y.}~\bibnamefont {Ouyang}},\ and\
  \bibinfo {author} {\bibfnamefont {J.}~\bibnamefont {Chen}},\ }\bibfield
  {title} {\bibinfo {title} {A perspective on the hydrodynamic phonon transport
  in two-dimensional materials},\ }\href {https://doi.org/10.1063/5.0056315}
  {\bibfield  {journal} {\bibinfo  {journal} {J. Appl. Phys.}\ }\textbf
  {\bibinfo {volume} {130}},\ \bibinfo {pages} {010902} (\bibinfo {year}
  {2021})}\BibitemShut {NoStop}%
\bibitem [{\citenamefont {Crossno}\ \emph {et~al.}(2016)\citenamefont
  {Crossno}, \citenamefont {Shi}, \citenamefont {Wang}, \citenamefont {Liu},
  \citenamefont {Harzheim}, \citenamefont {Lucas}, \citenamefont {Sachdev},
  \citenamefont {Kim}, \citenamefont {Taniguchi}, \citenamefont {Watanabe},
  \citenamefont {Ohki},\ and\ \citenamefont {Fong}}]{crossno_observation_2016}%
  \BibitemOpen
  \bibfield  {author} {\bibinfo {author} {\bibfnamefont {J.}~\bibnamefont
  {Crossno}}, \bibinfo {author} {\bibfnamefont {J.~K.}\ \bibnamefont {Shi}},
  \bibinfo {author} {\bibfnamefont {K.}~\bibnamefont {Wang}}, \bibinfo {author}
  {\bibfnamefont {X.}~\bibnamefont {Liu}}, \bibinfo {author} {\bibfnamefont
  {A.}~\bibnamefont {Harzheim}}, \bibinfo {author} {\bibfnamefont
  {A.}~\bibnamefont {Lucas}}, \bibinfo {author} {\bibfnamefont
  {S.}~\bibnamefont {Sachdev}}, \bibinfo {author} {\bibfnamefont
  {P.}~\bibnamefont {Kim}}, \bibinfo {author} {\bibfnamefont {T.}~\bibnamefont
  {Taniguchi}}, \bibinfo {author} {\bibfnamefont {K.}~\bibnamefont {Watanabe}},
  \bibinfo {author} {\bibfnamefont {T.~A.}\ \bibnamefont {Ohki}},\ and\
  \bibinfo {author} {\bibfnamefont {K.~C.}\ \bibnamefont {Fong}},\ }\bibfield
  {title} {\bibinfo {title} {Observation of the {Dirac} fluid and the breakdown
  of the {Wiedemann}-{Franz} law in graphene},\ }\href
  {https://doi.org/10.1126/science.aad0343} {\bibfield  {journal} {\bibinfo
  {journal} {Science}\ }\textbf {\bibinfo {volume} {351}},\ \bibinfo {pages}
  {1058} (\bibinfo {year} {2016})}\BibitemShut {NoStop}%
\bibitem [{\citenamefont {Bandurin}\ \emph {et~al.}(2016)\citenamefont
  {Bandurin}, \citenamefont {Torre}, \citenamefont {Kumar}, \citenamefont
  {Shalom}, \citenamefont {Tomadin}, \citenamefont {Principi}, \citenamefont
  {Auton}, \citenamefont {Khestanova}, \citenamefont {Novoselov}, \citenamefont
  {Grigorieva}, \citenamefont {Ponomarenko}, \citenamefont {Geim},\ and\
  \citenamefont {Polini}}]{bandurin_negative_2016}%
  \BibitemOpen
  \bibfield  {author} {\bibinfo {author} {\bibfnamefont {D.~A.}\ \bibnamefont
  {Bandurin}}, \bibinfo {author} {\bibfnamefont {I.}~\bibnamefont {Torre}},
  \bibinfo {author} {\bibfnamefont {R.~K.}\ \bibnamefont {Kumar}}, \bibinfo
  {author} {\bibfnamefont {M.~B.}\ \bibnamefont {Shalom}}, \bibinfo {author}
  {\bibfnamefont {A.}~\bibnamefont {Tomadin}}, \bibinfo {author} {\bibfnamefont
  {A.}~\bibnamefont {Principi}}, \bibinfo {author} {\bibfnamefont {G.~H.}\
  \bibnamefont {Auton}}, \bibinfo {author} {\bibfnamefont {E.}~\bibnamefont
  {Khestanova}}, \bibinfo {author} {\bibfnamefont {K.~S.}\ \bibnamefont
  {Novoselov}}, \bibinfo {author} {\bibfnamefont {I.~V.}\ \bibnamefont
  {Grigorieva}}, \bibinfo {author} {\bibfnamefont {L.~A.}\ \bibnamefont
  {Ponomarenko}}, \bibinfo {author} {\bibfnamefont {A.~K.}\ \bibnamefont
  {Geim}},\ and\ \bibinfo {author} {\bibfnamefont {M.}~\bibnamefont {Polini}},\
  }\bibfield  {title} {\bibinfo {title} {Negative local resistance caused by
  viscous electron backflow in graphene},\ }\href
  {https://doi.org/10.1126/science.aad0201} {\bibfield  {journal} {\bibinfo
  {journal} {Science}\ }\textbf {\bibinfo {volume} {351}},\ \bibinfo {pages}
  {1055} (\bibinfo {year} {2016})}\BibitemShut {NoStop}%
\bibitem [{\citenamefont {Moll}\ \emph {et~al.}(2016)\citenamefont {Moll},
  \citenamefont {Kushwaha}, \citenamefont {Nandi}, \citenamefont {Schmidt},\
  and\ \citenamefont {Mackenzie}}]{moll_evidence_2016}%
  \BibitemOpen
  \bibfield  {author} {\bibinfo {author} {\bibfnamefont {P.~J.~W.}\
  \bibnamefont {Moll}}, \bibinfo {author} {\bibfnamefont {P.}~\bibnamefont
  {Kushwaha}}, \bibinfo {author} {\bibfnamefont {N.}~\bibnamefont {Nandi}},
  \bibinfo {author} {\bibfnamefont {B.}~\bibnamefont {Schmidt}},\ and\ \bibinfo
  {author} {\bibfnamefont {A.~P.}\ \bibnamefont {Mackenzie}},\ }\bibfield
  {title} {\bibinfo {title} {Evidence for hydrodynamic electron flow in
  pdcoo2},\ }\href {https://doi.org/10.1126/science.aac8385} {\bibfield
  {journal} {\bibinfo  {journal} {Science}\ }\textbf {\bibinfo {volume}
  {351}},\ \bibinfo {pages} {1061} (\bibinfo {year} {2016})}\BibitemShut
  {NoStop}%
\bibitem [{\citenamefont {Sulpizio}\ \emph {et~al.}(2019)\citenamefont
  {Sulpizio}, \citenamefont {Ella}, \citenamefont {Rozen}, \citenamefont
  {Birkbeck}, \citenamefont {Perello}, \citenamefont {Dutta}, \citenamefont
  {Ben-Shalom}, \citenamefont {Taniguchi}, \citenamefont {Watanabe},
  \citenamefont {Holder}, \citenamefont {Queiroz}, \citenamefont {Principi},
  \citenamefont {Stern}, \citenamefont {Scaffidi}, \citenamefont {Geim},\ and\
  \citenamefont {Ilani}}]{sulpizio_visualizing_2019}%
  \BibitemOpen
  \bibfield  {author} {\bibinfo {author} {\bibfnamefont {J.~A.}\ \bibnamefont
  {Sulpizio}}, \bibinfo {author} {\bibfnamefont {L.}~\bibnamefont {Ella}},
  \bibinfo {author} {\bibfnamefont {A.}~\bibnamefont {Rozen}}, \bibinfo
  {author} {\bibfnamefont {J.}~\bibnamefont {Birkbeck}}, \bibinfo {author}
  {\bibfnamefont {D.~J.}\ \bibnamefont {Perello}}, \bibinfo {author}
  {\bibfnamefont {D.}~\bibnamefont {Dutta}}, \bibinfo {author} {\bibfnamefont
  {M.}~\bibnamefont {Ben-Shalom}}, \bibinfo {author} {\bibfnamefont
  {T.}~\bibnamefont {Taniguchi}}, \bibinfo {author} {\bibfnamefont
  {K.}~\bibnamefont {Watanabe}}, \bibinfo {author} {\bibfnamefont
  {T.}~\bibnamefont {Holder}}, \bibinfo {author} {\bibfnamefont
  {R.}~\bibnamefont {Queiroz}}, \bibinfo {author} {\bibfnamefont
  {A.}~\bibnamefont {Principi}}, \bibinfo {author} {\bibfnamefont
  {A.}~\bibnamefont {Stern}}, \bibinfo {author} {\bibfnamefont
  {T.}~\bibnamefont {Scaffidi}}, \bibinfo {author} {\bibfnamefont {A.~K.}\
  \bibnamefont {Geim}},\ and\ \bibinfo {author} {\bibfnamefont
  {S.}~\bibnamefont {Ilani}},\ }\bibfield  {title} {\bibinfo {title}
  {Visualizing {Poiseuille} flow of hydrodynamic electrons},\ }\href
  {https://doi.org/10.1038/s41586-019-1788-9} {\bibfield  {journal} {\bibinfo
  {journal} {Nature}\ }\textbf {\bibinfo {volume} {576}},\ \bibinfo {pages}
  {75} (\bibinfo {year} {2019})}\BibitemShut {NoStop}%
\bibitem [{\citenamefont {Gallagher}\ \emph {et~al.}(2019)\citenamefont
  {Gallagher}, \citenamefont {Yang}, \citenamefont {Lyu}, \citenamefont {Tian},
  \citenamefont {Kou}, \citenamefont {Zhang}, \citenamefont {Watanabe},
  \citenamefont {Taniguchi},\ and\ \citenamefont
  {Wang}}]{gallagher_quantum-critical_2019}%
  \BibitemOpen
  \bibfield  {author} {\bibinfo {author} {\bibfnamefont {P.}~\bibnamefont
  {Gallagher}}, \bibinfo {author} {\bibfnamefont {C.-S.}\ \bibnamefont {Yang}},
  \bibinfo {author} {\bibfnamefont {T.}~\bibnamefont {Lyu}}, \bibinfo {author}
  {\bibfnamefont {F.}~\bibnamefont {Tian}}, \bibinfo {author} {\bibfnamefont
  {R.}~\bibnamefont {Kou}}, \bibinfo {author} {\bibfnamefont {H.}~\bibnamefont
  {Zhang}}, \bibinfo {author} {\bibfnamefont {K.}~\bibnamefont {Watanabe}},
  \bibinfo {author} {\bibfnamefont {T.}~\bibnamefont {Taniguchi}},\ and\
  \bibinfo {author} {\bibfnamefont {F.}~\bibnamefont {Wang}},\ }\bibfield
  {title} {\bibinfo {title} {Quantum-critical conductivity of the {Dirac} fluid
  in graphene},\ }\href {https://doi.org/10.1126/science.aat8687} {\bibfield
  {journal} {\bibinfo  {journal} {Science}\ }\textbf {\bibinfo {volume}
  {364}},\ \bibinfo {pages} {158} (\bibinfo {year} {2019})}\BibitemShut
  {NoStop}%
\bibitem [{\citenamefont {Ella}\ \emph {et~al.}(2019)\citenamefont {Ella},
  \citenamefont {Rozen}, \citenamefont {Birkbeck}, \citenamefont {Ben-Shalom},
  \citenamefont {Perello}, \citenamefont {Zultak}, \citenamefont {Taniguchi},
  \citenamefont {Watanabe}, \citenamefont {Geim}, \citenamefont {Ilani},\ and\
  \citenamefont {Sulpizio}}]{ella_simultaneous_2019}%
  \BibitemOpen
  \bibfield  {author} {\bibinfo {author} {\bibfnamefont {L.}~\bibnamefont
  {Ella}}, \bibinfo {author} {\bibfnamefont {A.}~\bibnamefont {Rozen}},
  \bibinfo {author} {\bibfnamefont {J.}~\bibnamefont {Birkbeck}}, \bibinfo
  {author} {\bibfnamefont {M.}~\bibnamefont {Ben-Shalom}}, \bibinfo {author}
  {\bibfnamefont {D.}~\bibnamefont {Perello}}, \bibinfo {author} {\bibfnamefont
  {J.}~\bibnamefont {Zultak}}, \bibinfo {author} {\bibfnamefont
  {T.}~\bibnamefont {Taniguchi}}, \bibinfo {author} {\bibfnamefont
  {K.}~\bibnamefont {Watanabe}}, \bibinfo {author} {\bibfnamefont {A.~K.}\
  \bibnamefont {Geim}}, \bibinfo {author} {\bibfnamefont {S.}~\bibnamefont
  {Ilani}},\ and\ \bibinfo {author} {\bibfnamefont {J.~A.}\ \bibnamefont
  {Sulpizio}},\ }\bibfield  {title} {\bibinfo {title} {Simultaneous voltage and
  current density imaging of flowing electrons in two dimensions},\ }\href
  {https://doi.org/10.1038/s41565-019-0398-x} {\bibfield  {journal} {\bibinfo
  {journal} {Nat. Nanotechnol.}\ }\textbf {\bibinfo {volume} {14}},\ \bibinfo
  {pages} {480} (\bibinfo {year} {2019})}\BibitemShut {NoStop}%
\bibitem [{\citenamefont {Berdyugin}\ \emph {et~al.}(2019)\citenamefont
  {Berdyugin}, \citenamefont {Xu}, \citenamefont {Pellegrino}, \citenamefont
  {Kumar}, \citenamefont {Principi}, \citenamefont {Torre}, \citenamefont
  {Shalom}, \citenamefont {Taniguchi}, \citenamefont {Watanabe}, \citenamefont
  {Grigorieva}, \citenamefont {Polini}, \citenamefont {Geim},\ and\
  \citenamefont {Bandurin}}]{berdyugin_measuring_2019}%
  \BibitemOpen
  \bibfield  {author} {\bibinfo {author} {\bibfnamefont {A.~I.}\ \bibnamefont
  {Berdyugin}}, \bibinfo {author} {\bibfnamefont {S.~G.}\ \bibnamefont {Xu}},
  \bibinfo {author} {\bibfnamefont {F.~M.~D.}\ \bibnamefont {Pellegrino}},
  \bibinfo {author} {\bibfnamefont {R.~K.}\ \bibnamefont {Kumar}}, \bibinfo
  {author} {\bibfnamefont {A.}~\bibnamefont {Principi}}, \bibinfo {author}
  {\bibfnamefont {I.}~\bibnamefont {Torre}}, \bibinfo {author} {\bibfnamefont
  {M.~B.}\ \bibnamefont {Shalom}}, \bibinfo {author} {\bibfnamefont
  {T.}~\bibnamefont {Taniguchi}}, \bibinfo {author} {\bibfnamefont
  {K.}~\bibnamefont {Watanabe}}, \bibinfo {author} {\bibfnamefont {I.~V.}\
  \bibnamefont {Grigorieva}}, \bibinfo {author} {\bibfnamefont
  {M.}~\bibnamefont {Polini}}, \bibinfo {author} {\bibfnamefont {A.~K.}\
  \bibnamefont {Geim}},\ and\ \bibinfo {author} {\bibfnamefont {D.~A.}\
  \bibnamefont {Bandurin}},\ }\bibfield  {title} {\bibinfo {title} {Measuring
  hall viscosity of graphene’s electron fluid},\ }\href
  {https://doi.org/10.1126/science.aau0685} {\bibfield  {journal} {\bibinfo
  {journal} {Science}\ }\textbf {\bibinfo {volume} {364}},\ \bibinfo {pages}
  {162} (\bibinfo {year} {2019})}\BibitemShut {NoStop}%
\bibitem [{\citenamefont {Levchenko}\ and\ \citenamefont
  {Schmalian}(2020)}]{levchenko_transport_2020}%
  \BibitemOpen
  \bibfield  {author} {\bibinfo {author} {\bibfnamefont {A.}~\bibnamefont
  {Levchenko}}\ and\ \bibinfo {author} {\bibfnamefont {J.}~\bibnamefont
  {Schmalian}},\ }\bibfield  {title} {\bibinfo {title} {Transport properties of
  strongly coupled electron–phonon liquids},\ }\href
  {https://doi.org/10.1016/j.aop.2020.168218} {\bibfield  {journal} {\bibinfo
  {journal} {Ann. Phys.}\ }\textbf {\bibinfo {volume} {419}},\ \bibinfo {pages}
  {168218} (\bibinfo {year} {2020})}\BibitemShut {NoStop}%
\bibitem [{\citenamefont {Huang}\ and\ \citenamefont
  {Lucas}(2021)}]{huang_electron-phonon_2021}%
  \BibitemOpen
  \bibfield  {author} {\bibinfo {author} {\bibfnamefont {X.}~\bibnamefont
  {Huang}}\ and\ \bibinfo {author} {\bibfnamefont {A.}~\bibnamefont {Lucas}},\
  }\bibfield  {title} {\bibinfo {title} {Electron-phonon hydrodynamics},\
  }\href {https://doi.org/10.1103/PhysRevB.103.155128} {\bibfield  {journal}
  {\bibinfo  {journal} {Phys. Rev. B}\ }\textbf {\bibinfo {volume} {103}},\
  \bibinfo {pages} {155128} (\bibinfo {year} {2021})}\BibitemShut {NoStop}%
\bibitem [{\citenamefont {Narozhny}(2019)}]{narozhny_electronic_2019}%
  \BibitemOpen
  \bibfield  {author} {\bibinfo {author} {\bibfnamefont {B.~N.}\ \bibnamefont
  {Narozhny}},\ }\bibfield  {title} {\bibinfo {title} {Electronic hydrodynamics
  in graphene},\ }\href {https://doi.org/10.1016/j.aop.2019.167979} {\bibfield
  {journal} {\bibinfo  {journal} {Ann. Phys.}\ }\textbf {\bibinfo {volume}
  {411}},\ \bibinfo {pages} {167979} (\bibinfo {year} {2019})}\BibitemShut
  {NoStop}%
\bibitem [{\citenamefont {Hardy}\ and\ \citenamefont
  {Jaswal}(1971)}]{hardy_velocity_1971}%
  \BibitemOpen
  \bibfield  {author} {\bibinfo {author} {\bibfnamefont {R.~J.}\ \bibnamefont
  {Hardy}}\ and\ \bibinfo {author} {\bibfnamefont {S.~S.}\ \bibnamefont
  {Jaswal}},\ }\bibfield  {title} {\bibinfo {title} {Velocity of {Second}
  {Sound} in {NaF}},\ }\href {https://doi.org/10.1103/PhysRevB.3.4385}
  {\bibfield  {journal} {\bibinfo  {journal} {Phys. Rev. B}\ }\textbf {\bibinfo
  {volume} {3}},\ \bibinfo {pages} {4385} (\bibinfo {year} {1971})}\BibitemShut
  {NoStop}%
\bibitem [{\citenamefont {Beardo}\ \emph {et~al.}(2021)\citenamefont {Beardo},
  \citenamefont {L{\'o}pez-Su{\'a}rez}, \citenamefont {P{\'e}rez},
  \citenamefont {Sendra}, \citenamefont {Alonso}, \citenamefont {Melis},
  \citenamefont {Bafaluy}, \citenamefont {Camacho}, \citenamefont {Colombo},
  \citenamefont {Rurali}, \citenamefont {Alvarez},\ and\ \citenamefont
  {Reparaz}}]{beardo_observation_2021}%
  \BibitemOpen
  \bibfield  {author} {\bibinfo {author} {\bibfnamefont {A.}~\bibnamefont
  {Beardo}}, \bibinfo {author} {\bibfnamefont {M.}~\bibnamefont
  {L{\'o}pez-Su{\'a}rez}}, \bibinfo {author} {\bibfnamefont {L.~A.}\
  \bibnamefont {P{\'e}rez}}, \bibinfo {author} {\bibfnamefont {L.}~\bibnamefont
  {Sendra}}, \bibinfo {author} {\bibfnamefont {M.~I.}\ \bibnamefont {Alonso}},
  \bibinfo {author} {\bibfnamefont {C.}~\bibnamefont {Melis}}, \bibinfo
  {author} {\bibfnamefont {J.}~\bibnamefont {Bafaluy}}, \bibinfo {author}
  {\bibfnamefont {J.}~\bibnamefont {Camacho}}, \bibinfo {author} {\bibfnamefont
  {L.}~\bibnamefont {Colombo}}, \bibinfo {author} {\bibfnamefont
  {R.}~\bibnamefont {Rurali}}, \bibinfo {author} {\bibfnamefont {F.~X.}\
  \bibnamefont {Alvarez}},\ and\ \bibinfo {author} {\bibfnamefont {J.~S.}\
  \bibnamefont {Reparaz}},\ }\bibfield  {title} {\bibinfo {title} {Observation
  of second sound in a rapidly varying temperature field in ge},\ }\bibfield
  {journal} {\bibinfo  {journal} {Sci. Adv}\ }\textbf {\bibinfo {volume} {7}},\
  \href {https://doi.org/10.1126/sciadv.abg4677} {10.1126/sciadv.abg4677}
  (\bibinfo {year} {2021})\BibitemShut {NoStop}%
\bibitem [{\citenamefont {Simoncelli}\ \emph {et~al.}(2020)\citenamefont
  {Simoncelli}, \citenamefont {Marzari},\ and\ \citenamefont
  {Cepellotti}}]{simoncelli_generalization_2020}%
  \BibitemOpen
  \bibfield  {author} {\bibinfo {author} {\bibfnamefont {M.}~\bibnamefont
  {Simoncelli}}, \bibinfo {author} {\bibfnamefont {N.}~\bibnamefont
  {Marzari}},\ and\ \bibinfo {author} {\bibfnamefont {A.}~\bibnamefont
  {Cepellotti}},\ }\bibfield  {title} {\bibinfo {title} {Generalization of
  {Fourier}'s {Law} into {Viscous} {Heat} {Equations}},\ }\href
  {https://doi.org/10.1103/PhysRevX.10.011019} {\bibfield  {journal} {\bibinfo
  {journal} {Phys. Rev. X}\ }\textbf {\bibinfo {volume} {10}},\ \bibinfo
  {pages} {011019} (\bibinfo {year} {2020})}\BibitemShut {NoStop}%
\bibitem [{\citenamefont {Callaway}(1959)}]{callaway_model_1959}%
  \BibitemOpen
  \bibfield  {author} {\bibinfo {author} {\bibfnamefont {J.}~\bibnamefont
  {Callaway}},\ }\bibfield  {title} {\bibinfo {title} {Model for {Lattice}
  {Thermal} {Conductivity} at {Low} {Temperatures}},\ }\href
  {https://doi.org/10.1103/PhysRev.113.1046} {\bibfield  {journal} {\bibinfo
  {journal} {Phys. Rev.}\ }\textbf {\bibinfo {volume} {113}},\ \bibinfo {pages}
  {1046} (\bibinfo {year} {1959})}\BibitemShut {NoStop}%
\bibitem [{\citenamefont {Guo}\ and\ \citenamefont
  {Wang}(2015)}]{guo_phonon_2015}%
  \BibitemOpen
  \bibfield  {author} {\bibinfo {author} {\bibfnamefont {Y.}~\bibnamefont
  {Guo}}\ and\ \bibinfo {author} {\bibfnamefont {M.}~\bibnamefont {Wang}},\
  }\bibfield  {title} {\bibinfo {title} {Phonon hydrodynamics and its
  applications in nanoscale heat transport},\ }\href
  {https://doi.org/https://doi.org/10.1016/j.physrep.2015.07.003} {\bibfield
  {journal} {\bibinfo  {journal} {Phys. Rep.}\ }\textbf {\bibinfo {volume}
  {595}},\ \bibinfo {pages} {1 } (\bibinfo {year} {2015})}\BibitemShut
  {NoStop}%
\bibitem [{Note1()}]{Note1}%
  \BibitemOpen
  \bibinfo {note} {These expressions hold in the constant relaxation time
  approximation, i.e., $\langle \tau _c \rangle _\kappa = \langle \tau _c
  \rangle _q=\langle \tau _c \rangle $.}\BibitemShut {Stop}%
\bibitem [{\citenamefont {Landau}\ and\ \citenamefont
  {Lifshitz}(1987)}]{LandauBook6}%
  \BibitemOpen
  \bibfield  {author} {\bibinfo {author} {\bibfnamefont {L.~D.}\ \bibnamefont
  {Landau}}\ and\ \bibinfo {author} {\bibfnamefont {E.~M.}\ \bibnamefont
  {Lifshitz}},\ }\href@noop {} {\emph {\bibinfo {title} {Fluid Mechanics}}},\
  \bibinfo {edition} {second edition}\ ed.\ (\bibinfo  {publisher} {Pergamon
  Press},\ \bibinfo {year} {1987})\BibitemShut {NoStop}%
\bibitem [{\citenamefont {Volz}(2001)}]{volz_thermal_2001}%
  \BibitemOpen
  \bibfield  {author} {\bibinfo {author} {\bibfnamefont {S.~G.}\ \bibnamefont
  {Volz}},\ }\bibfield  {title} {\bibinfo {title} {Thermal insulating behavior
  in crystals at high frequencies},\ }\href
  {https://doi.org/10.1103/PhysRevLett.87.074301} {\bibfield  {journal}
  {\bibinfo  {journal} {Phys. Rev. Lett.}\ }\textbf {\bibinfo {volume} {87}},\
  \bibinfo {pages} {074301} (\bibinfo {year} {2001})}\BibitemShut {NoStop}%
\bibitem [{\citenamefont {Chaput}(2013)}]{chaput_direct_2013}%
  \BibitemOpen
  \bibfield  {author} {\bibinfo {author} {\bibfnamefont {L.}~\bibnamefont
  {Chaput}},\ }\bibfield  {title} {\bibinfo {title} {Direct solution to the
  linearized phonon boltzmann equation},\ }\href
  {https://doi.org/10.1103/PhysRevLett.110.265506} {\bibfield  {journal}
  {\bibinfo  {journal} {Phys. Rev. Lett.}\ }\textbf {\bibinfo {volume} {110}},\
  \bibinfo {pages} {265506} (\bibinfo {year} {2013})}\BibitemShut {NoStop}%
\bibitem [{\citenamefont {Hua}\ and\ \citenamefont
  {Lindsay}(2020)}]{hua_space_2020}%
  \BibitemOpen
  \bibfield  {author} {\bibinfo {author} {\bibfnamefont {C.}~\bibnamefont
  {Hua}}\ and\ \bibinfo {author} {\bibfnamefont {L.}~\bibnamefont {Lindsay}},\
  }\bibfield  {title} {\bibinfo {title} {Space-time dependent thermal
  conductivity in nonlocal thermal transport},\ }\href
  {https://doi.org/10.1103/PhysRevB.102.104310} {\bibfield  {journal} {\bibinfo
   {journal} {Phys. Rev. B}\ }\textbf {\bibinfo {volume} {102}},\ \bibinfo
  {pages} {104310} (\bibinfo {year} {2020})}\BibitemShut {NoStop}%
\bibitem [{\citenamefont {Koh}\ and\ \citenamefont
  {Cahill}(2007)}]{koh_frequency_2007}%
  \BibitemOpen
  \bibfield  {author} {\bibinfo {author} {\bibfnamefont {Y.~K.}\ \bibnamefont
  {Koh}}\ and\ \bibinfo {author} {\bibfnamefont {D.~G.}\ \bibnamefont
  {Cahill}},\ }\bibfield  {title} {\bibinfo {title} {Frequency dependence of
  the thermal conductivity of semiconductor alloys},\ }\href
  {https://doi.org/10.1103/PhysRevB.76.075207} {\bibfield  {journal} {\bibinfo
  {journal} {Phys. Rev. B}\ }\textbf {\bibinfo {volume} {76}},\ \bibinfo
  {pages} {075207} (\bibinfo {year} {2007})}\BibitemShut {NoStop}%
\bibitem [{\citenamefont {Mariani}\ and\ \citenamefont {von
  Oppen}(2008)}]{Mariani_flexural_2008}%
  \BibitemOpen
  \bibfield  {author} {\bibinfo {author} {\bibfnamefont {E.}~\bibnamefont
  {Mariani}}\ and\ \bibinfo {author} {\bibfnamefont {F.}~\bibnamefont {von
  Oppen}},\ }\bibfield  {title} {\bibinfo {title} {Flexural phonons in
  free-standing graphene},\ }\href
  {https://doi.org/10.1103/PhysRevLett.100.076801} {\bibfield  {journal}
  {\bibinfo  {journal} {Phys. Rev. Lett.}\ }\textbf {\bibinfo {volume} {100}},\
  \bibinfo {pages} {076801} (\bibinfo {year} {2008})}\BibitemShut {NoStop}%
\bibitem [{\citenamefont {Aseginolaza}\ \emph {et~al.}(2020)\citenamefont
  {Aseginolaza}, \citenamefont {Cea}, \citenamefont {Bianco}, \citenamefont
  {Monacelli}, \citenamefont {Calandra}, \citenamefont {Bergara}, \citenamefont
  {Mauri},\ and\ \citenamefont {Errea}}]{aseginolaza_bending_2020}%
  \BibitemOpen
  \bibfield  {author} {\bibinfo {author} {\bibfnamefont {U.}~\bibnamefont
  {Aseginolaza}}, \bibinfo {author} {\bibfnamefont {T.}~\bibnamefont {Cea}},
  \bibinfo {author} {\bibfnamefont {R.}~\bibnamefont {Bianco}}, \bibinfo
  {author} {\bibfnamefont {L.}~\bibnamefont {Monacelli}}, \bibinfo {author}
  {\bibfnamefont {M.}~\bibnamefont {Calandra}}, \bibinfo {author}
  {\bibfnamefont {A.}~\bibnamefont {Bergara}}, \bibinfo {author} {\bibfnamefont
  {F.}~\bibnamefont {Mauri}},\ and\ \bibinfo {author} {\bibfnamefont
  {I.}~\bibnamefont {Errea}},\ }\href@noop {} {} (\bibinfo {year} {2020}),\
  \Eprint {https://arxiv.org/abs/2005.12047} {arXiv:2005.12047
  [cond-mat.mes-hall]} \BibitemShut {NoStop}%
\bibitem [{\citenamefont {Carrete}\ \emph {et~al.}(2016)\citenamefont
  {Carrete}, \citenamefont {Li}, \citenamefont {Lindsay}, \citenamefont
  {Broido}, \citenamefont {Gallego},\ and\ \citenamefont
  {Mingo}}]{Jes2016Physically}%
  \BibitemOpen
  \bibfield  {author} {\bibinfo {author} {\bibfnamefont {J.}~\bibnamefont
  {Carrete}}, \bibinfo {author} {\bibfnamefont {W.}~\bibnamefont {Li}},
  \bibinfo {author} {\bibfnamefont {L.}~\bibnamefont {Lindsay}}, \bibinfo
  {author} {\bibfnamefont {D.~A.}\ \bibnamefont {Broido}}, \bibinfo {author}
  {\bibfnamefont {L.~J.}\ \bibnamefont {Gallego}},\ and\ \bibinfo {author}
  {\bibfnamefont {N.}~\bibnamefont {Mingo}},\ }\bibfield  {title} {\bibinfo
  {title} {Physically founded phonon dispersions of few-layer materials and the
  case of borophene},\ }\href {https://doi.org/10.1080/21663831.2016.1174163}
  {\bibfield  {journal} {\bibinfo  {journal} {Mater. Res. Lett.}\ }\textbf
  {\bibinfo {volume} {4}},\ \bibinfo {pages} {204} (\bibinfo {year}
  {2016})}\BibitemShut {NoStop}%
\bibitem [{\citenamefont {Zhang}\ \emph {et~al.}(2020)\citenamefont {Zhang},
  \citenamefont {Ouyang}, \citenamefont {Guo}, \citenamefont {Nakayama},
  \citenamefont {Nomura}, \citenamefont {Volz},\ and\ \citenamefont
  {Chen}}]{zhang_hydrodynamic_2020}%
  \BibitemOpen
  \bibfield  {author} {\bibinfo {author} {\bibfnamefont {Z.}~\bibnamefont
  {Zhang}}, \bibinfo {author} {\bibfnamefont {Y.}~\bibnamefont {Ouyang}},
  \bibinfo {author} {\bibfnamefont {Y.}~\bibnamefont {Guo}}, \bibinfo {author}
  {\bibfnamefont {T.}~\bibnamefont {Nakayama}}, \bibinfo {author}
  {\bibfnamefont {M.}~\bibnamefont {Nomura}}, \bibinfo {author} {\bibfnamefont
  {S.}~\bibnamefont {Volz}},\ and\ \bibinfo {author} {\bibfnamefont
  {J.}~\bibnamefont {Chen}},\ }\bibfield  {title} {\bibinfo {title}
  {Hydrodynamic phonon transport in bulk crystalline polymers},\ }\href
  {https://doi.org/10.1103/PhysRevB.102.195302} {\bibfield  {journal} {\bibinfo
   {journal} {Phys. Rev. B}\ }\textbf {\bibinfo {volume} {102}},\ \bibinfo
  {pages} {195302} (\bibinfo {year} {2020})}\BibitemShut {NoStop}%
\bibitem [{\citenamefont {Pereira}\ and\ \citenamefont
  {Donadio}(2013)}]{Pereira13}%
  \BibitemOpen
  \bibfield  {author} {\bibinfo {author} {\bibfnamefont {L.~F.~C.}\
  \bibnamefont {Pereira}}\ and\ \bibinfo {author} {\bibfnamefont
  {D.}~\bibnamefont {Donadio}},\ }\bibfield  {title} {\bibinfo {title}
  {Divergence of the thermal conductivity in uniaxially strained graphene},\
  }\href {https://doi.org/10.1103/PhysRevB.87.125424} {\bibfield  {journal}
  {\bibinfo  {journal} {Phys. Rev. B}\ }\textbf {\bibinfo {volume} {87}},\
  \bibinfo {pages} {125424} (\bibinfo {year} {2013})}\BibitemShut {NoStop}%
\bibitem [{\citenamefont {Bonini}\ \emph {et~al.}(2012)\citenamefont {Bonini},
  \citenamefont {Garg},\ and\ \citenamefont {Marzari}}]{bonini_acoustic_2012}%
  \BibitemOpen
  \bibfield  {author} {\bibinfo {author} {\bibfnamefont {N.}~\bibnamefont
  {Bonini}}, \bibinfo {author} {\bibfnamefont {J.}~\bibnamefont {Garg}},\ and\
  \bibinfo {author} {\bibfnamefont {N.}~\bibnamefont {Marzari}},\ }\bibfield
  {title} {\bibinfo {title} {Acoustic {Phonon} {Lifetimes} and {Thermal}
  {Transport} in {Free}-{Standing} and {Strained} {Graphene}},\ }\href
  {https://doi.org/10.1021/nl202694m} {\bibfield  {journal} {\bibinfo
  {journal} {Nano Lett.}\ }\textbf {\bibinfo {volume} {12}},\ \bibinfo {pages}
  {2673} (\bibinfo {year} {2012})}\BibitemShut {NoStop}%
\bibitem [{\citenamefont {Lindsay}\ \emph {et~al.}(2014)\citenamefont
  {Lindsay}, \citenamefont {Li}, \citenamefont {Carrete}, \citenamefont
  {Mingo}, \citenamefont {Broido},\ and\ \citenamefont
  {Reinecke}}]{lindsay_phonon_2014}%
  \BibitemOpen
  \bibfield  {author} {\bibinfo {author} {\bibfnamefont {L.}~\bibnamefont
  {Lindsay}}, \bibinfo {author} {\bibfnamefont {W.}~\bibnamefont {Li}},
  \bibinfo {author} {\bibfnamefont {J.}~\bibnamefont {Carrete}}, \bibinfo
  {author} {\bibfnamefont {N.}~\bibnamefont {Mingo}}, \bibinfo {author}
  {\bibfnamefont {D.~A.}\ \bibnamefont {Broido}},\ and\ \bibinfo {author}
  {\bibfnamefont {T.~L.}\ \bibnamefont {Reinecke}},\ }\bibfield  {title}
  {\bibinfo {title} {Phonon thermal transport in strained and unstrained
  graphene from first principles},\ }\href
  {https://doi.org/10.1103/PhysRevB.89.155426} {\bibfield  {journal} {\bibinfo
  {journal} {Phys. Rev. B}\ }\textbf {\bibinfo {volume} {89}},\ \bibinfo
  {pages} {155426} (\bibinfo {year} {2014})}\BibitemShut {NoStop}%
\bibitem [{\citenamefont {Fugallo}\ \emph {et~al.}(2014)\citenamefont
  {Fugallo}, \citenamefont {Cepellotti}, \citenamefont {Paulatto},
  \citenamefont {Lazzeri}, \citenamefont {Marzari},\ and\ \citenamefont
  {Mauri}}]{fugallo_thermal_2014}%
  \BibitemOpen
  \bibfield  {author} {\bibinfo {author} {\bibfnamefont {G.}~\bibnamefont
  {Fugallo}}, \bibinfo {author} {\bibfnamefont {A.}~\bibnamefont {Cepellotti}},
  \bibinfo {author} {\bibfnamefont {L.}~\bibnamefont {Paulatto}}, \bibinfo
  {author} {\bibfnamefont {M.}~\bibnamefont {Lazzeri}}, \bibinfo {author}
  {\bibfnamefont {N.}~\bibnamefont {Marzari}},\ and\ \bibinfo {author}
  {\bibfnamefont {F.}~\bibnamefont {Mauri}},\ }\bibfield  {title} {\bibinfo
  {title} {Thermal {Conductivity} of {Graphene} and {Graphite}: {Collective}
  {Excitations} and {Mean} {Free} {Paths}},\ }\href
  {https://doi.org/10.1021/nl502059f} {\bibfield  {journal} {\bibinfo
  {journal} {Nano Lett.}\ }\textbf {\bibinfo {volume} {14}},\ \bibinfo {pages}
  {6109} (\bibinfo {year} {2014})}\BibitemShut {NoStop}%
\bibitem [{\citenamefont {Kuang}\ \emph {et~al.}(2015)\citenamefont {Kuang},
  \citenamefont {Lindsay},\ and\ \citenamefont {Huang}}]{kuang_unusual_2015}%
  \BibitemOpen
  \bibfield  {author} {\bibinfo {author} {\bibfnamefont {Y.}~\bibnamefont
  {Kuang}}, \bibinfo {author} {\bibfnamefont {L.}~\bibnamefont {Lindsay}},\
  and\ \bibinfo {author} {\bibfnamefont {B.}~\bibnamefont {Huang}},\ }\bibfield
   {title} {\bibinfo {title} {Unusual {Enhancement} in {Intrinsic} {Thermal}
  {Conductivity} of {Multilayer} {Graphene} by {Tensile} {Strains}},\ }\href
  {https://doi.org/10.1021/acs.nanolett.5b02403} {\bibfield  {journal}
  {\bibinfo  {journal} {Nano Lett.}\ }\textbf {\bibinfo {volume} {15}},\
  \bibinfo {pages} {6121} (\bibinfo {year} {2015})}\BibitemShut {NoStop}%
\bibitem [{\citenamefont {Kuang}\ \emph {et~al.}(2016)\citenamefont {Kuang},
  \citenamefont {Lindsay}, \citenamefont {Shi}, \citenamefont {Wang},\ and\
  \citenamefont {Huang}}]{kuang_thermal_2016}%
  \BibitemOpen
  \bibfield  {author} {\bibinfo {author} {\bibfnamefont {Y.}~\bibnamefont
  {Kuang}}, \bibinfo {author} {\bibfnamefont {L.}~\bibnamefont {Lindsay}},
  \bibinfo {author} {\bibfnamefont {S.}~\bibnamefont {Shi}}, \bibinfo {author}
  {\bibfnamefont {X.}~\bibnamefont {Wang}},\ and\ \bibinfo {author}
  {\bibfnamefont {B.}~\bibnamefont {Huang}},\ }\bibfield  {title} {\bibinfo
  {title} {Thermal conductivity of graphene mediated by strain and size},\
  }\href {https://doi.org/10.1016/j.ijheatmasstransfer.2016.05.072} {\bibfield
  {journal} {\bibinfo  {journal} {Int. J. Heat Mass Transf.}\ }\textbf
  {\bibinfo {volume} {101}},\ \bibinfo {pages} {772} (\bibinfo {year}
  {2016})}\BibitemShut {NoStop}%
\bibitem [{\citenamefont {Gu}\ and\ \citenamefont
  {Yang}(2015)}]{gu_first-principles_2015}%
  \BibitemOpen
  \bibfield  {author} {\bibinfo {author} {\bibfnamefont {X.}~\bibnamefont
  {Gu}}\ and\ \bibinfo {author} {\bibfnamefont {R.}~\bibnamefont {Yang}},\
  }\bibfield  {title} {\bibinfo {title} {First-principles prediction of
  phononic thermal conductivity of silicene: {A} comparison with graphene},\
  }\href {https://doi.org/10.1063/1.4905540} {\bibfield  {journal} {\bibinfo
  {journal} {J. Appl. Phys.}\ }\textbf {\bibinfo {volume} {117}},\ \bibinfo
  {pages} {025102} (\bibinfo {year} {2015})}\BibitemShut {NoStop}%
\bibitem [{\citenamefont {{Xu}}\ \emph {et~al.}(2014)\citenamefont {{Xu}},
  \citenamefont {{Pereira}}, \citenamefont {{Wang}}, \citenamefont {{Wu}},
  \citenamefont {{Zhang}}, \citenamefont {{Zhao}}, \citenamefont {{Bae}},
  \citenamefont {{Tinh Bui}}, \citenamefont {{Xie}}, \citenamefont {{Thong}},
  \citenamefont {{Hong}}, \citenamefont {{Loh}}, \citenamefont {{Donadio}},
  \citenamefont {{Li}},\ and\ \citenamefont {{{\"O}zyilmaz}}}]{Xu2014}%
  \BibitemOpen
  \bibfield  {author} {\bibinfo {author} {\bibfnamefont {X.}~\bibnamefont
  {{Xu}}}, \bibinfo {author} {\bibfnamefont {L.~F.~C.}\ \bibnamefont
  {{Pereira}}}, \bibinfo {author} {\bibfnamefont {Y.}~\bibnamefont {{Wang}}},
  \bibinfo {author} {\bibfnamefont {J.}~\bibnamefont {{Wu}}}, \bibinfo {author}
  {\bibfnamefont {K.}~\bibnamefont {{Zhang}}}, \bibinfo {author} {\bibfnamefont
  {X.}~\bibnamefont {{Zhao}}}, \bibinfo {author} {\bibfnamefont
  {S.}~\bibnamefont {{Bae}}}, \bibinfo {author} {\bibfnamefont
  {C.}~\bibnamefont {{Tinh Bui}}}, \bibinfo {author} {\bibfnamefont
  {R.}~\bibnamefont {{Xie}}}, \bibinfo {author} {\bibfnamefont {J.~T.~L.}\
  \bibnamefont {{Thong}}}, \bibinfo {author} {\bibfnamefont {B.~H.}\
  \bibnamefont {{Hong}}}, \bibinfo {author} {\bibfnamefont {K.~P.}\
  \bibnamefont {{Loh}}}, \bibinfo {author} {\bibfnamefont {D.}~\bibnamefont
  {{Donadio}}}, \bibinfo {author} {\bibfnamefont {B.}~\bibnamefont {{Li}}},\
  and\ \bibinfo {author} {\bibfnamefont {B.}~\bibnamefont {{{\"O}zyilmaz}}},\
  }\bibfield  {title} {\bibinfo {title} {Length-dependent thermal conductivity
  in suspended single-layer graphene},\ }\href
  {https://doi.org/10.1038/ncomms4689} {\bibfield  {journal} {\bibinfo
  {journal} {Nat. Commun.}\ }\textbf {\bibinfo {volume} {5}},\ \bibinfo {pages}
  {3689} (\bibinfo {year} {2014})}\BibitemShut {NoStop}%
\bibitem [{\citenamefont {Landau}(1941)}]{landau1941two}%
  \BibitemOpen
  \bibfield  {author} {\bibinfo {author} {\bibfnamefont {L.}~\bibnamefont
  {Landau}},\ }\bibfield  {title} {\bibinfo {title} {Two-fluid model of liquid
  helium ii},\ }\href@noop {} {\bibfield  {journal} {\bibinfo  {journal} {J.
  Phys. Ussr}\ }\textbf {\bibinfo {volume} {5}},\ \bibinfo {pages} {71}
  (\bibinfo {year} {1941})}\BibitemShut {NoStop}%
\bibitem [{\citenamefont {Kresse}\ and\ \citenamefont
  {Furthm\"uller}(1996)}]{PhysRevB.54.11169}%
  \BibitemOpen
  \bibfield  {author} {\bibinfo {author} {\bibfnamefont {G.}~\bibnamefont
  {Kresse}}\ and\ \bibinfo {author} {\bibfnamefont {J.}~\bibnamefont
  {Furthm\"uller}},\ }\bibfield  {title} {\bibinfo {title} {Efficient iterative
  schemes for ab initio total-energy calculations using a plane-wave basis
  set},\ }\href {https://doi.org/10.1103/PhysRevB.54.11169} {\bibfield
  {journal} {\bibinfo  {journal} {Phys. Rev. B}\ }\textbf {\bibinfo {volume}
  {54}},\ \bibinfo {pages} {11169} (\bibinfo {year} {1996})}\BibitemShut
  {NoStop}%
\bibitem [{\citenamefont {Kresse}\ and\ \citenamefont
  {Furthm{\"u}ller}(1996)}]{KRESSE199615}%
  \BibitemOpen
  \bibfield  {author} {\bibinfo {author} {\bibfnamefont {G.}~\bibnamefont
  {Kresse}}\ and\ \bibinfo {author} {\bibfnamefont {J.}~\bibnamefont
  {Furthm{\"u}ller}},\ }\bibfield  {title} {\bibinfo {title} {Efficiency of
  ab-initio total energy calculations for metals and semiconductors using a
  plane-wave basis set},\ }\href
  {https://doi.org/https://doi.org/10.1016/0927-0256(96)00008-0} {\bibfield
  {journal} {\bibinfo  {journal} {Comput. Mater. Sci.}\ }\textbf {\bibinfo
  {volume} {6}},\ \bibinfo {pages} {15} (\bibinfo {year} {1996})}\BibitemShut
  {NoStop}%
\bibitem [{\citenamefont {Togo}\ and\ \citenamefont {Tanaka}(2015)}]{phonopy}%
  \BibitemOpen
  \bibfield  {author} {\bibinfo {author} {\bibfnamefont {A.}~\bibnamefont
  {Togo}}\ and\ \bibinfo {author} {\bibfnamefont {I.}~\bibnamefont {Tanaka}},\
  }\bibfield  {title} {\bibinfo {title} {First principles phonon calculations
  in materials science},\ }\href
  {https://doi.org/https://doi.org/10.1016/j.scriptamat.2015.07.021} {\bibfield
   {journal} {\bibinfo  {journal} {Scr. Mater.}\ }\textbf {\bibinfo {volume}
  {108}},\ \bibinfo {pages} {1} (\bibinfo {year} {2015})}\BibitemShut {NoStop}%
\bibitem [{\citenamefont {Togo}\ \emph {et~al.}(2015)\citenamefont {Togo},
  \citenamefont {Chaput},\ and\ \citenamefont {Tanaka}}]{phono3py}%
  \BibitemOpen
  \bibfield  {author} {\bibinfo {author} {\bibfnamefont {A.}~\bibnamefont
  {Togo}}, \bibinfo {author} {\bibfnamefont {L.}~\bibnamefont {Chaput}},\ and\
  \bibinfo {author} {\bibfnamefont {I.}~\bibnamefont {Tanaka}},\ }\bibfield
  {title} {\bibinfo {title} {Distributions of phonon lifetimes in brillouin
  zones},\ }\href {https://doi.org/10.1103/PhysRevB.91.094306} {\bibfield
  {journal} {\bibinfo  {journal} {Phys. Rev. B}\ }\textbf {\bibinfo {volume}
  {91}},\ \bibinfo {pages} {094306} (\bibinfo {year} {2015})}\BibitemShut
  {NoStop}%
\end{thebibliography}%

\end{document}